\documentclass[a4j]{article}
\usepackage[dvipdfm]{graphicx}
\usepackage{bm}
\usepackage{amsmath}
\usepackage{mathrsfs}
\usepackage{ascmac}
\usepackage{amsthm}
\usepackage{amssymb}
\usepackage{feynmp}
\usepackage{url}
\usepackage{setspace}
\usepackage{cite}
\makeatletter

\newcommand{\figcaption}[1]{\def\@captype{figure}\caption{#1}}
\newcommand{\tblcaption}[1]{\def\@captype{table}\caption{#1}}

\makeatother

\newcommand{\beq}{\begin{equation}}
\newcommand{\eeq}{\end{equation}}
\newcommand{\beqa}{\begin{eqnarray}}
\newcommand{\eeqa}{\end{eqnarray}}
\newcommand{\Tr}{\text{Tr}}

\newcommand{\bra}[1]{\left \langle #1 \right |}
\newcommand{\ket}[1]{\left | #1 \right \rangle}
\newcommand{\av}[1]{\left\langle #1 \right\rangle}
\newcommand{\braket}[2]{\left. \left\langle #1 \right| #2 \right \rangle} 
 
\newcommand{\kket}[1]{| #1 \hspace{-.9mm}\left. \right>}
\newcommand{\bbra}[1]{  \hspace{-.6mm}\left<\hspace{-.4mm} \right. #1 |  }
\newcommand{\aav}[1]{ \hspace{-.6mm}\left<\hspace{-.4mm} \right. #1 \hspace{-.9mm}\left. \right> }

\newcommand{\ii}{\mathrm{i}}
\newcommand{\ee}{\mathrm{e}}
\newcommand{\dd}{\mathrm{d}}

\newcommand{\bz}{k_{\text{B}}}

\begin{document}

\title{Quantum nonequilibrium equalities with absolute irreversibility}

\author{Ken Funo$^1$, Y\^{u}to Murashita$^1$ and Masahito Ueda$^{1,2}$ \\ $^1$ Department of Physics, The University of Tokyo,\\  7-3-1 Hongo, Bunkyo-ku, Tokyo 113-0033, Japan\\ $^2$ Center for Emergent Matter Science (CEMS), RIKEN, Wako, Saitama 351-0198, Japan}

\maketitle


\begin{abstract}
We derive quantum nonequilibrium equalities in absolutely irreversible processes. Here by absolute irreversibility we mean that in the backward process the density matrix does not return to the subspace spanned by those eigenvectors that have nonzero weight in the initial density matrix. Since the initial state of a memory and the postmeasurement state of the system are usually restricted to a subspace, absolute irreversibility occurs during the measurement and feedback processes. An additional entropy produced in absolute irreversible processes needs to be taken into account to derive nonequilibrium equalities. We discuss a model of a feedback control on a qubit system to illustrate the obtained equalities. By introducing $N$ heat baths each composed of a qubit and letting them interact with the system, we show how the entropy reduction via feedback control can be converted into work. An explicit form of extractable work in the presence of absolute irreversibility is given.
\end{abstract}

\section{Introduction}

\begin{figure}[h,t]
\begin{center}
\includegraphics[width=.4\textwidth]{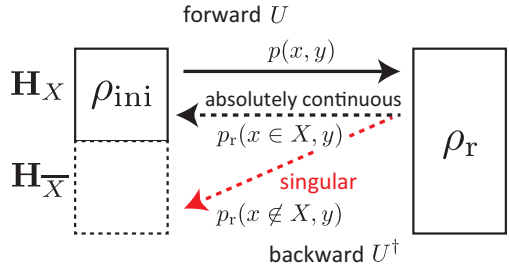}
\caption{\label{fig:abld} Schematic illustration of an absolutely irreversible process. We start from an initial state described by the density matrix $\rho_{\mathrm{ini}}=\sum_{x}p_{\mathrm{ini}}(x)\kket{\psi(x)}\bbra{\psi(x)}$, where its support is restricted to the subspace ${\bf H}_{X}$. Here the label $X$ is a set of variables $x$ satisfying $p_{\mathrm{ini}}(x)\neq 0$; thus the subspace ${\bf H}_{X}$ is spanned by the set of orthonormal states $\{\psi(x)\}_{x\in X}$. We denote ${\bf H}_{\bar{X}}$ as the orthogonal compliment of ${\bf H}_{X}$. We consider the case in which the forward process is given by the time evolution via a unitary operator $U$. We assume that the initial state of the backward process described by the density matrix $\rho_{\mathrm{r}}=\sum_{y}p_{\mathrm{r}}(y)\kket{\phi(y)}\bbra{\phi(y)}$. The backward process is described by the time reversal of the forward process via the unitary operator $U^{\dagger}$. Then, with nonzero probability, the density matrix of the backward protocol evolves in time into the space outside of the subspace ${\bf H}_{X}$. In terms of path probabilities, the forward and backward probabilities are given by $p(x,y)$ and $p_{\mathrm{r}}(x,y)$, respectively. By using Lebesgue's decomposition theorem~\cite{Lebesgue1,Lebesgue2}, we can uniquely decompose the backward probability into two parts: $p_{\mathrm{r}}(x\in X,y)$ and $p_{\mathrm{r}}(x\not\in X,y)$ which are absolutely continuous and singular with respect to $p(x,y)$, respectively. We call such a process absolutely irreversible in the sense that there is no one-to-one correspondence between the forward and backward probabilities, and that the entropy production diverges for the singular part, i.e., $\sigma(x\not\in X,y)=\ln p(x\not\in X,y)-\ln p_{\mathrm{r}}(x\not\in X,y)=-\infty$.}
\end{center}
\end{figure}

Nonequilibrium equalities~\cite{Evans1,Evans,Gallavotti,Crooks,Jarzynski1,Jarzynski2,Seifert,Campisi3,Tasaki,Kurchan,Esposito,Monnai,characteristic1,Talkner3,characteristic2,Campisi,fluctuation1,Horowitz1,Horowitz2,Chetrite,Hu,Crooks2} such as fluctuation theorems and Jarzynski equalities have attracted a great deal of interest in the field of nonequilibrium statistical mechanics. They give general insights into thermodynamic quantities in nonequilibrium processes irrespective of details of individual systems. For example, the Jarzynski equality relates work done on the system in a nonequilbrium process to the equilibrium free-energy difference. Nonequilibrium equalities have been obtained in both classical~\cite{Evans,Evans1,Gallavotti,Crooks,Jarzynski1,Jarzynski2,Seifert} and  quantum~\cite{Tasaki,Kurchan,Esposito,Monnai,characteristic1,characteristic2,Campisi3,Campisi,fluctuation1,Talkner3,Horowitz1,Horowitz2,Chetrite,Hu,Crooks2} systems, and generalized to situations involving feedback control~\cite{SagawaJar,Sagawa3,Sagawa4,Morikuni,Funo}. Due to recent advancement in experimental techniques, nonequilibrium equalities have been experimentally verified for classical systems such as a single-molecule RNA~\cite{Jarexperiment,Flucex}, and they have been vindicated in a quantum regime  using a trapped ion system~\cite{qJarexperiment}. Also, feedback control on a Brownian particle was carried out to experimentally demonstrate Maxwell's demon~\cite{Maxwell}, and the generalized Jarzynski equality for a feedback-controlled system was verified~\cite{Toyabe}.

It is known that the Jarzynski equalities are inapplicable to such cases as free expansion~\cite{Jarin4,Jarin5,Jarin1,Jarin2,Jarin3} and feedback control involving projective measurements~\cite{Morikuni} because in these cases there exist those forward paths with vanishing probability that have the corresponsing backward paths with nonzero probabilities. We shall call such processes absolutely irreversible. We give examples of absolutely irreversible processes in Sec.~\ref{sec:example}. Recently, nonequilibrium equalities were obtained that can be applied to absolutely irreversible processes, including the processes mentioned above~\cite{Murashita,Ashida}. We extend this idea to quantum systems and derive quantum fluctuation theorems and Jarzynski equalities with absolute irreversibility. For the quantum case, absolute irreversibility occurs when the initial state $\rho_{\mathrm{ini}}=\sum_{x}p_{\mathrm{ini}}(x)\kket{\psi(x)}\bbra{\psi(x)}$ (with $p_{\mathrm{ini}}(x)\neq 0$ for $x\in X$) is restricted to the subspace ${\bf H}_{X}$ (spanned by $\{\kket{\psi(x)}\}_{x\in X}$) of the total Hilbert space, and the density matrix of the backward process is not confined to that subspace, i.e., $\lambda=\sum_{x\not\in X}\bra{\psi(x)}\tilde{\rho}\ket{\psi(x)}\neq 0$, where $\tilde{\rho}$ is the final density matrix of the backward process (see Fig.~\ref{fig:abld}). Then, the initially localized state expands into a larger space, as happens in free expansion, with the probability $\lambda$. Absolute irreversibility is likely to occur in measurement and feedback processes since the initial state of the memory and the postmeasurement state are localized in general, and the projective measurement on the memory and the effect of (inefficient) feedback control let these states expand to a space larger than the subspace of the initial state, resulting in additional entropy production. By subtracting the absolutely irreversible part in a mathematically well-defined manner, we derive those nonequilibrium equalities for measurement and feedback processes which give stronger restrictions on entropy productions or work compared with previously known results~\cite{Sagawa1,Sagawa2,Funo}. We note that  a quantum Jarzynski equality under feedback control with projective measurement was obtained in Ref.~\cite{Morikuni}, where the issue of absolute irreversibility was circumvented by introducing classical errors on measurement outcomes.

This paper is organized as follows. In Sec.~\ref{sec:qfint}, we derive nonequilibrium equalities without feedback control for quantum systems. We introduce the concept of absolute irreversibility and discuss how the nonequilibrium equalities are modified by this effect. In Sec.~\ref{sec:main}, we derive nonequilibrium equalities with feedback control in the presence  of absolute irreversibility during feedback control and the measurement process. In Sec.~\ref{sec:example}, we give an example of the quantum piston to analyze the free expansion of the gas with absolute irreversibility. We also give an example of the feedback control on a qubit system to illustrate our work. In Sec.~\ref{sec:conclusion}, we summarize the main results of this paper.

\section{\label{sec:qfint} Nonequilibrium equalities without feedback control}
\subsection{Setup}
Let the initial state of the system be $\rho_{\mathrm{ini}}$ and let the system evolve in time according to a unitary evolution: 
\beq
U=\text{T}\exp\left(-\ii\int_{0}^{t_{\mathrm{fin}}}H(t)\dd t\right),\label{unitaryevo}
\eeq
where $H(t)$ is the time-dependent Hamiltonian. The final state of the system is given by
\beq
\rho_{\mathrm{fin}}=U\rho_{\mathrm{ini}} U^{\dagger}. \label{finalunitary}
\eeq 

We define the entropy production, which measures the irreversibility of the process, as
\beq
\aav{\sigma}=-S(\rho_{\mathrm{ini}})-\Tr[\rho_{\mathrm{fin}}\ln \rho_{\mathrm{r}}], \label{entdef}
\eeq
where $S(\rho_{\mathrm{ini}})=-\Tr[\rho_{\mathrm{ini}}\ln\rho_{\mathrm{ini}}]$ is the von Neumann entropy and $\rho_{\mathrm{r}}$ is a reference state which can be chosen arbitrarily~\cite{Sagawa5}. Because of Eq.~(\ref{finalunitary}), the entropy production defined here is nothing but the quantum relative entropy between the final state and the reference state:
\beq
\aav{\sigma}=-S(\rho_{\mathrm{fin}})-\Tr[\rho_{\mathrm{fin}}\ln\rho_{\mathrm{r}}]=S(\rho_{\mathrm{fin}}||\rho_{\mathrm{r}})\geq 0, \label{slineq}
\eeq
where the inequality results from the nonnegativity of the quantum relative entropy~\cite{Nielsen}. Different choices of the reference states lead to different entropy productions~\cite{Seifert}. Here we give two examples.

\subsubsection{Examples of the choice of reference states and the corresponding entropy productions}
\paragraph{1. Dissipated work}
We defined the dissipated work $W_{\mathrm{d}}$ in terms of the work $\av{W}$ done by the system (or work that can be extracted from the system)\footnote{We keep this sign convention for work throughout this paper .} and the equilibrium free energy difference $\Delta F=F_{\mathrm{fin}}-F_{\mathrm{ini}}$ as
\beq
W_{\mathrm{d}}=-\beta \left\langle W\right\rangle -\beta\Delta F, \label{dissipatedwork}
\eeq
and assume the initial state to be the canonical distribution
\beq
\rho_{\mathrm{ini}}=\ee^{-\beta(H_{\mathrm{ini}}-F_{\mathrm{ini}})}, \label{disassupa}
\eeq
where $H_{\mathrm{ini}}=H(0)$. If we choose the reference state to be the canonical distribution with respect to the final Hamiltonian $H_{\mathrm{fin}}=H(t_{\mathrm{fin}})$
\beq
\rho_{\mathrm{r}}=\ee^{-\beta(H_{\mathrm{fin}}-F_{\mathrm{fin}})}, \label{disassupb}
\eeq
then Eq.~(\ref{entdef}) reduces to the dissipated work
\beqa
\aav{\sigma}=-\beta(\aav{W}+\Delta F)=W_{\mathrm{d}}\geq 0, \label{disswork}
\eeqa
where we define work during the nonequilibrium process by the energy change of the system:
\beq
\left\langle W\right\rangle =\Tr[H_{\mathrm{ini}}\rho_{\mathrm{ini}}]-\Tr[H_{\mathrm{fin}}\rho_{\mathrm{fin}}].
\eeq
The above argument applies to an isolated system. In the presence of a heat bath, the total Hamiltonian in Eq.~(\ref{unitaryevo}) is given by
\beq
H(t)=H^{S}(t)+V^{SB}(t)+H^{B}, \label{hamiltonian}
\eeq
where $H^{S}(t)$ is the system Hamiltonian depending on time via external control parameters, $H^{B}$ is the Hamiltonian of the heat bath, and $V^{SB}(t)$ is the interaction between the system and the heat bath. In Eq.~(\ref{hamiltonian}), we assume that the interaction $V^{SB}(t)$ is either turned off at the initial and final states, i.e., 
\beq
V^{SB}(0)=V^{SB}(t_{\mathrm{fin}})=0, \label{intassump}
\eeq
or the interaction is independent of time and very weak, i.e.,
\beq
V^{SB}(t)=\kappa V^{SB} ,\hspace{5mm} \kappa<<1. \label{intassumpb}
\eeq
Later, we discuss the validity of these assumption we made for this system-heat bath interaction. We also use the abbreviations $H^{S}_{\mathrm{ini}}=H^{S}(0)$ and $H^{S}_{\mathrm{fin}}=H^{S}(t_{\mathrm{fin}})$. We first consider the case~(\ref{intassump}). Then
\beq
\rho_{\mathrm{ini}}=\ee^{-\beta(H_{\mathrm{ini}}^{S}-F_{\mathrm{ini}}^{S})}\otimes\ee^{-\beta(H^{B}-F^{B})}, \label{dissbatha}
\eeq
holds and the choice of reference state in Eq.~(\ref{disassupb}) leads to
\beq
\rho_{\mathrm{r}}=\ee^{-\beta(H_{\mathrm{fin}}^{S}-F_{\mathrm{fin}}^{S})}\otimes\ee^{-\beta(H^{B}-F^{B})}. \label{dissbathb}
\eeq
Combining Eqs.~(\ref{entdef}), (\ref{dissbatha}) and (\ref{dissbathb}), we reproduce Eq.~(\ref{disswork}):
\beq
\aav{\sigma}=-\beta(\aav{W}+\Delta F^{S})=W_{\mathrm{d}}\geq 0, \label{dissworkbath}
\eeq
where the work appearing in Eq.~(\ref{dissworkbath}) is given by
\beq
\aav{W}=\Tr[H^{S}_{\mathrm{ini}}\rho^{S}_{\mathrm{ini}}]-\Tr[H^{S}_{\mathrm{fin}}\rho^{S}_{\mathrm{fin}}]+ Q. \label{sltotentpro}
\eeq
Here, we interpret the heat $Q$ as the energy that is transfered from the heat bath to the system during the process:
\beq
Q=\Tr[H^{B}\rho^{B}_{\mathrm{ini}}]-\Tr[H^{B}\rho^{B}_{\mathrm{fin}}]. \label{heatdef}
\eeq

Now let us consider the validity of the assumption~(\ref{intassump}) we made for the interaction $V(t)$. We have in mind a system that is attached to the heat bath at the initial time and detached at the final time. One example of a system satisfying this condition is a cavity field interacting with a sequence of atoms passing through the cavity, where atoms can act as a heat bath to the cavity field~\cite{cavity}. There are some subtlety for the definition of work in this case because there might be a contribution to work~(\ref{sltotentpro}) from the action of switching on and off the interaction. We can avoid this subtlety by assuming that the interaction is turned on and off adiabatically. We also note that we can adopt the framework of continuous measurement (by considering many heat baths interacting with a system one by one and) by taking the limit in which the interaction time with each environment is sufficiently small, but the coupling strength is assumed to scale as the inverse of the square root of the interaction time. In this limit, the stochastic master equation was derived in Ref.\cite{Horowitz2}, and fluctuation theorems were obtained.

Next, let us consider the case where we assume Eq.~(\ref{intassumpb}) for the interaction $V(t)$. We note that when the interaction is always present, the initial state has a correlation between the system and the heat bath. However, in the weak coupling limit, Eq.~(\ref{dissbatha}) is correct up to the second order of the coupling strength:
\beq
\rho_{\mathrm{ini}}=\ee^{-\beta(H_{\mathrm{ini}}-F_{\mathrm{ini}})}=\ee^{-\beta(H_{\mathrm{ini}}^{S}-F_{\mathrm{ini}}^{S})}\otimes\ee^{-\beta(H^{B}-F^{B})}+O(\kappa^{2}). 
\eeq
When we use the definition of work and heat as given in Eqs.~(\ref{sltotentpro}) and (\ref{heatdef}), we must assume that the energy change due to the interaction energy is small. We assume that the total energy change is divided into the energy change of the system and that of the heat bath for weak coupling as discussed in Ref.~\cite{Talkner3}:
\beq
\Tr[H_{\mathrm{ini}}\rho_{\mathrm{ini}}]-\Tr[H_{\mathrm{fin}}\rho_{\mathrm{fin}}]\simeq(\Tr[H^{S}_{\mathrm{ini}}\rho^{S}_{\mathrm{ini}}]-\Tr[H^{S}_{\mathrm{fin}}\rho^{S}_{\mathrm{fin}}])+(\Tr[H^{B}\rho^{B}_{\mathrm{ini}}]-\Tr[H^{B}\rho^{B}_{\mathrm{fin}}]) .
\eeq
The main results of the rest of this paper is based on the assumption~(\ref{intassump}) for the interaction, but the same result can be derived if we assume~(\ref{intassumpb}) instead of~(\ref{intassump}) (in particular, a quantum Jarzynski equality without absolute irreversibility was derived in Ref.~\cite{Talkner3} for the weak coupling interaction.)

\paragraph{2. Total entropy production}
To relate the entropy production to the total entropy production, we consider a system composed of a system and a heat bath, and use the same Hamiltonian as in Eq.~(\ref{hamiltonian}). We assume that the initial state of the heat bath is given by the canonical distribution
\beq
\rho^{SB}_{\mathrm{ini}}=\rho^{S}_{\mathrm{ini}}\otimes\ee^{-\beta(H^{B}-F^{B})}, \label{totentproa}
\eeq
and choose the reference state as follows:
\beq
\rho^{SB}_{\mathrm{r}}=\rho^{S}_{\mathrm{fin}}\otimes\ee^{-\beta(H^{B}-F^{B})}. \label{totentprob}
\eeq
Combining Eqs.~(\ref{entdef}), (\ref{totentproa}) and (\ref{totentprob}), we obtain
\beq
\aav{\sigma}=\Delta S-\beta Q=\sigma_{\mathrm{tot}}\geq 0, \label{totentprosig}
\eeq
where $\Delta S=S(\rho^{S}_{\mathrm{fin}})-S(\rho^{S}_{\mathrm{ini}})$ is a change in the von Neumann entropy of the system and $Q$ is the heat defined in Eq.~(\ref{heatdef}). If we interpret heat as the entropy produced in the heat bath, Eq.~(\ref{totentprosig}) expresses the entropy that is produced for the total system during the protocol; $\sigma_{\mathrm{tot}}$ is therefore called the total entropy production.

Equation~(\ref{slineq}) leads to second-law-like inequalities for entropy productions, e.g., for dissipated work and total entropy production, and the nonnegativity of the entropy production shows that there is dissipation in a given process~\cite{Sagawa5}. The process is thermodynamically reversible if and only if the equality in (\ref{slineq}) holds (for example, if the dissipated work $W_{\mathrm{d}}$ or the total entropy production $\sigma_{\mathrm{tot}}$ vanishes).

\subsection{Quantum fluctuation theorem}
Next, we derive quantum fluctuation theorems by expressing the initial state in the diagonal basis.m and the initial state as  $\rho_{\mathrm{ini}}=\sum_{x}p_{\mathrm{ini}}(x)\kket{\psi(x)}\bbra{\psi(x)}$, where $\{\kket{\psi(x)}\}$ is an orthonormal basis set, and the reference state as $\rho_{\mathrm{r}}=\sum_{y}p_{\mathrm{r}}(y)\kket{\phi(y)}\bbra{\phi(y)}$. The entropy production can then be calculated as
\beqa
\aav{\sigma}&=&\sum_{x}p_{\mathrm{ini}}(x)\ln p_{\mathrm{ini}}(x)-\Tr\biggl[\rho_{\mathrm{fin}}\ln \biggl(\sum_{y}p_{\mathrm{r}}(y)\kket{\phi(y)}\bbra{\phi(y)}\biggr)\biggr] \nonumber \\
&=&\sum_{x}p_{\mathrm{ini}}(x)\ln p_{\mathrm{ini}}(x)-\sum_{y}\bbra{\phi(y)}\rho_{\mathrm{fin}}\kket{\phi(y)}\ln p_{\mathrm{r}}(y) \nonumber \\
&=& \sum_{x}p_{\mathrm{ini}}(x)\ln p_{\mathrm{ini}}(x)-\sum_{x,y}p(x,y)\ln p_{\mathrm{r}}(y) \nonumber \\
&=&\sum_{x,y}p(x,y)\ln\frac{p_{\mathrm{ini}}(x)}{p_{\mathrm{r}}(y)}, \label{sigmadec}
\eeqa
where
\beq
p(x,y)=p_{\mathrm{ini}}(x)p(y|x)
\eeq
and
\beq
p(y|x)=|\bbra{\phi(y)}U\kket{\psi(x)}|^{2}
\eeq
is the transition probability from the state $\ket{\psi(x)}$ to $\ket{\phi(y)}$ via the unitary operator $U$. Such a transition is characterized by a set of labels $(x,y)$. In deriving the third line in Eq.~(\ref{sigmadec}), we used the relation
\beqa
\bbra{\phi(y)}\rho_{\mathrm{fin}}\kket{\phi(y)}&=&\bbra{\phi(y)}U\rho_{\mathrm{ini}}U^{\dagger}\kket{\phi(y)} \nonumber \\
&=&\sum_{x}p_{\mathrm{ini}}(x)|\bbra{\phi(y)}U\kket{\psi(x)}|^{2}.
\eeqa
From Eq.~(\ref{sigmadec}), we define the following unaveraged entropy production:
\beq
\sigma(x,y)=\ln \frac{p_{\mathrm{ini}}(x)}{p_{\mathrm{r}}(y)}. \label{entpropath}
\eeq
Next, we introduce the reference probability distribution
\beq
p_{\mathrm{r}}(x,y)=p_{\mathrm{r}}(y)\tilde{p}(x|y), \label{refprob}
\eeq
where
\beq
\tilde{p}(x|y)=|\bbra{\psi(x)}U^{\dagger}\kket{\phi(y)}|^{2}=p(y|x) \label{transitionrev}
\eeq
is the transition probability from $\ket{\phi(y)}$ to $\ket{\psi(x)}$ via $U^{\dagger}$. Equation (\ref{refprob}) gives the probability of the backward process that starts from the reference state and evolves in time via $U^{\dagger}$. It follows from Eq.~(\ref{transitionrev}) that the entropy production is expressed in terms of the forward and reference probabilities as follows:
\beq
\sigma(x,y)=\ln \frac{p_{\mathrm{ini}}(x)p(y|x)}{p_{\mathrm{r}}(y)\tilde{p}(x|y)}=\ln\frac{p(x,y)}{p_{\mathrm{r}}(x,y)}. \label{entratio}
\eeq
Now we derive the quantum fluctuation theorem by using the above definition of entropy production~(\ref{entpropath}). Since the sum of reference probability is unity, we have
\beq
\sum_{x,y}p_{\mathrm{r}}(x,y)=\Tr[U^{\dagger}\rho_{\mathrm{r}}U]=\Tr[\rho_{\mathrm{r}}]=1.
\eeq
The entropy production is given by the ratio between the forward and reference probabilities. However, if the forward probability vanishes and the corresponding reference probability does not, the logarithm of the ratio in Eq.~(\ref{entratio}) diverges. To deal with such singular situations, we divide the reference probability into two parts:
\beq
1=\sum_{x,y}p_{\mathrm{r}}(x,y)=\sum_{x\in X,y}p_{\mathrm{r}}(x,y)+\sum_{x\not\in X,y}p_{\mathrm{r}}(x,y) \label{refprobdec},
\eeq
where $X=\{x|p_{\mathrm{i}}(x)\neq 0\}$. Since we can take the ratio between the forward and reference probabilities for the first term of the right-hand side of Eq.~(\ref{refprobdec}), we have
\beqa
1&=&\sum_{x\in X,y}\frac{p_{\mathrm{r}}(x,y)}{p(x,y)}p(x,y)+\lambda  \nonumber \\
&=&\sum_{x\in X,y}p(x,y)\ee^{-\sigma(x,y)}+\lambda \nonumber \\
&=&\aav{\ee^{-\sigma}}+\lambda, \label{devfluc}
\eeqa
where
\beq
\lambda=\sum_{x\not\in X,y}p_{\mathrm{r}}(x,y) \label{singular}
\eeq
gives the total probability of those backward processes that do not return to the subspace spanned by $\{\ket{\psi(x)}\}_{x\in X}$. In an ordinary irreversible process, the process is stochastically reversible in the sense that the backward path returns to the initial state with nonzero probability since there is a one-to-one correspondence between the forward and backward paths, i.e., the entropy production is finite~\footnote{To be precise, when $p_{\mathrm{r}}(x,y)=0\ \land\ p(x,y)\neq 0$, the entropy production in Eq.~(\ref{entratio}) positively diverges. However, this does not cause any problem because $e^{-\sigma(x,y)}=p_{\mathrm{r}}(x,y)/p(x,y)=0$ remains finite and so is the first term on the left-hand side of Eq.~(\ref{devfluc}).} for all $(x,y)$ in Eq.~(\ref{entratio}). However, the path labeled by the set of variables $(x\not\in X,y)$ is not even stochastically reversible since the formal definition of the entropy production negatively diverges, i.e., 
\beq
\sigma(x\not\in X,y)=\ln\frac{0}{p_{\mathrm{r}}(x,y)}=-\infty,
\eeq
and we call this type of irreversibility absolute irreversibility~\cite{Murashita}. A schematic illustration of an absolutely irreversible process is shown in Fig.~\ref{fig:abld}.

By rewriting Eq.~(\ref{devfluc}), we obtain a quantum fluctuation theorem in the presence of absolute irreversibility:
\beq
\aav{\ee^{-\sigma}}=1-\lambda .\label{Qfluc}
\eeq
By using the Jensen's inequality, i.e., $\aav{\ee^{x}}\geq \ee^{\av{x}}$, we obtain the following inequality for the entropy production:
\beq
\av{\sigma}\geq -\ln(1-\lambda). \label{Entpro}
\eeq
This result shows that in the presence of absolute irreversibility the entropy production must be positive and not less than $-\ln(1-\lambda)> 0$, giving a stronger constraint compared with the second law-like inequality $\av{\sigma}\geq 0$. Note that only when there is no absolute irreversibility, i.e., $\lambda=0$, the conventional fluctuation theorem is reproduced: $\aav{\ee^{-\sigma}}=1$.

In the classical case, a decomposition similar to Eq.~(\ref{refprobdec}) can be carried out in a general framework using the probability measure~\cite{Murashita}. To see this, let us denote the forward and reference probability measures in phase space as $\mathcal{M}$ and $\mathcal{M}^{\mathrm{r}}$, respectively. According to Lebesgue's decomposition theorem~\cite{Lebesgue1,Lebesgue2}, $\mathcal{M}^{\mathrm{r}}$ is uniquely decomposed into two parts: $\mathcal{M}^{\mathrm{r}}=\mathcal{M}^{\mathrm{r}}_{\mathrm{AC}}+\mathcal{M}^{\mathrm{r}}_{\mathrm{S}}$, where $\mathcal{M}^{\mathrm{r}}_{\mathrm{AC}}$ and $\mathcal{M}^{\mathrm{r}}_{\mathrm{S}}$ are absolutely continuous and singular with respect to $\mathcal{M}$, respectively. Provided that the probability distribution of a quantum process in this setup is labeled by discrete variables, the decomposition of the reference probability is carried out by dividing variables into two parts: the variables with nonvanishing forward probabilities ($x\in X$) and the variables with vanishing forward probabilities ($x\not\in X$). Then, $\mathcal{M}^{\mathrm{r}}_{\mathrm{AC}}$ corresponds to $p^{\mathrm{r}}(x\in X,y)$ and $\mathcal{M}^{\mathrm{r}}_{\mathrm{S}}$ corresponds to $p^{\mathrm{r}}(x\not\in X,y)$, and this decomposition is unique as ensured by Lebesgue's decomposition theorem.

Note that the absence of absolute irreversibility and the requirement for the ``ergodic consistency" discussed in Ref.~\cite{Evans} are different concepts. The ergodic consistency requires that for all initial phase space $\Gamma(0)$ with nonzero probability $f(\Gamma(0),0)\neq 0$, the corresponding initial probability distribution of the time-reversed process is nonzero , i.e., $f(\Gamma^{\dagger}(t),0)\neq 0$, where $f(\Gamma,s)$ is the probability distribution of the system in the phase space point $\Gamma$ at time $s$. Therefore, the ergodic consistency requires that for all nonzero forward path probabilities, the corresponding reference (backward) path probabilities are nonzero. In contrast,  the absence of absolute irreversibility requires that for all nonzero reference (backward) path probabilities, the corresponding forward path probabilities are nonzero. The two conditions are different in the sense that in the former case, the backward protocol (especially, the initial state of the backward process) is fixed and thus the condition on the forward path probability is imposed, whereas and in the latter case, the forward protocol is fixed and thus the condition on the backward path probability is imposed.

\subsection{Quantum Jarzynski equality}
We now derive the quantum Jarzynski equality by assuming that the initial state is given by the canonical distribution~(\ref{dissbatha}) and by taking the reference state as given in Eq.~(\ref{dissbathb}). For convenience, we use the notation $x=(x_{1},x_{2})$ and $y=(y_{1},y_{2})$, where the subscript $1$ refers to the system and $2$ to the heat bath. By assumption, we have
\beq
\kket{\psi(x)}=\kket{E_{\mathrm{ini}}^{S}(x_{1})}\otimes\kket{E^{B}(x_{2})},\hspace{5mm} \kket{\phi(y)}=\kket{E^{S}_{\mathrm{fin}}(y_{1})}\otimes\kket{E^{B}(y_{2})},
\eeq
where $\kket{E_{\mathrm{ini}}^{S}(x_{1})}$ and $\kket{E^{S}_{\mathrm{fin}}(y_{1})}$ are energy eigenstates of the initial and final Hamiltonians of the system, respectively, and $\kket{E^{B}(x_{2})}$ is the energy eigenstate of the heat bath. Now the unaveraged entropy production~(\ref{entpropath}) is related to work by
\beq
\sigma(x,y)=-\beta(W(x,y)+\Delta F^{S}), \label{defunaventpro}
\eeq
where
\beqa
W(x,y)&=&E^{S}_{\mathrm{ini}}(x_{1})-E^{S}_{\mathrm{fin}}(y_{1})+E^{B}(x_{2})-E^{B}(y_{2}) \notag \\
&=&E^{S}_{\mathrm{ini}}(x_{1})-E^{S}_{\mathrm{fin}}(y_{1})+\beta Q(x_{2},y_{2}),
\eeqa
is the unaveraged work done by the system and $Q(x_{2},y_{2})$ is the unaveraged heat flowing into the system. 

Substituting Eq.~(\ref{defunaventpro}) into Eq.~(\ref{Qfluc}), we obtain the quantum Jarzynski equality in the presence of absolute irreversibility:
\beq
\aav{\ee^{-\beta(W+\Delta F)}}=1-\lambda. \label{Jarflucresult}
\eeq
Substituting Eq.~(\ref{defunaventpro}) into Eq.~(\ref{Entpro}), we obtain the second-law like inequality
\beq
\av{W}\leq -\Delta F^{S}+k_{\mathrm{B}}T\ln(1-\lambda).
\eeq
Since the canonical distribution is full rank, i.e.,
\beq
p^{S}_{\mathrm{ini}}(x_{1})=\ee^{-\beta(E^{S}_{\mathrm{ini}}(x_{1})-F^{S}_{\mathrm{ini}})}\neq 0 \hspace{5mm}\text{for all }x_{1},
\eeq
there is no absolute irreversibility, i.e., $\lambda=0$. However, if we prepare the initial state in a local equilibrium state, there is a possibility that the process is absolutely irreversible and the effect of nonzero $\lambda$ restricts the extractable work. For simplicity, let us divide the Hamiltonian of the system into two parts $H_{\mathrm{ini}}^{S}=H_{\mathrm{ini}}^{S_{X}}\oplus H_{\mathrm{ini}}^{S_{\bar{X}}}$ and prepare the initial state as the canonical distribution that is restricted to the subspace corresponding to $H^{S_{X}}_{\mathrm{ini}}$:
\beq
\rho^{S}_{\mathrm{ini}}=\ee^{-\beta(H^{S_{X}}_{\mathrm{ini}}-F_{\mathrm{ini}}^{S_{X}})}=\sum_{x_{1}\in X}\ee^{-\beta(E^{S}_{\mathrm{ini}}(x_{1})-F_{\mathrm{ini}}^{S_{X}})}\kket{E_{\mathrm{ini}}^{S}(x_{1})}\bbra{E_{\mathrm{ini}}^{S}(x_{1})},
\eeq
where $\{\kket{E_{\mathrm{ini}}^{S}(x_{1})}\}_{x_{1}\in X}$ is the energy eigenstate of the Hamiltonian $H^{S_{X}}_{\mathrm{ini}}$. Then, $\{\kket{E_{\mathrm{ini}}^{S}(x_{1})}\}_{x_{1}\not\in X}$ is the energy eigenstate of the Hamiltonian $H^{S_{\bar{X}}}_{\mathrm{ini}}$. Now $\lambda$ is given by the total probability of the backward process that the system returns to the subspace spanned by $\{\kket{E_{\mathrm{ini}}^{S}(x_{1})}\}_{x_{1}\not\in X}$:
\beq
\lambda=\sum_{x_{1}\not\in X} \left\langle E_{\mathrm{ini}}^{S}(x_{1})\right| \Tr_{B}\left[ U^{\dagger SB}(\rho^{S}_{\mathrm{can,fin}}\otimes\rho^{B}_{\mathrm{can}})U^{SB}\right]  \left| E_{\mathrm{ini}}^{S}(x_{1})\right\rangle,
\eeq
where $\rho^{S}_{\mathrm{can,fin}}\otimes\rho^{B}_{\mathrm{can}}$ is given by the right-hand side of Eq.~(\ref{dissbathb}).
When the initially localized state expands into the total Hilbert space, the process would be absolutely irreversible and a positive entropy is produced during this process. The effect of absolute irreversibility is to lower the extractable work by $k_{\mathrm{B}}T|\ln(1-\lambda)|$.

\section{\label{sec:main} Nonequilibrium equalities with feedback control}
\subsection{Formulation of the problem}

\begin{figure}[h,t]
\begin{center}
\includegraphics[width=.5\textwidth]{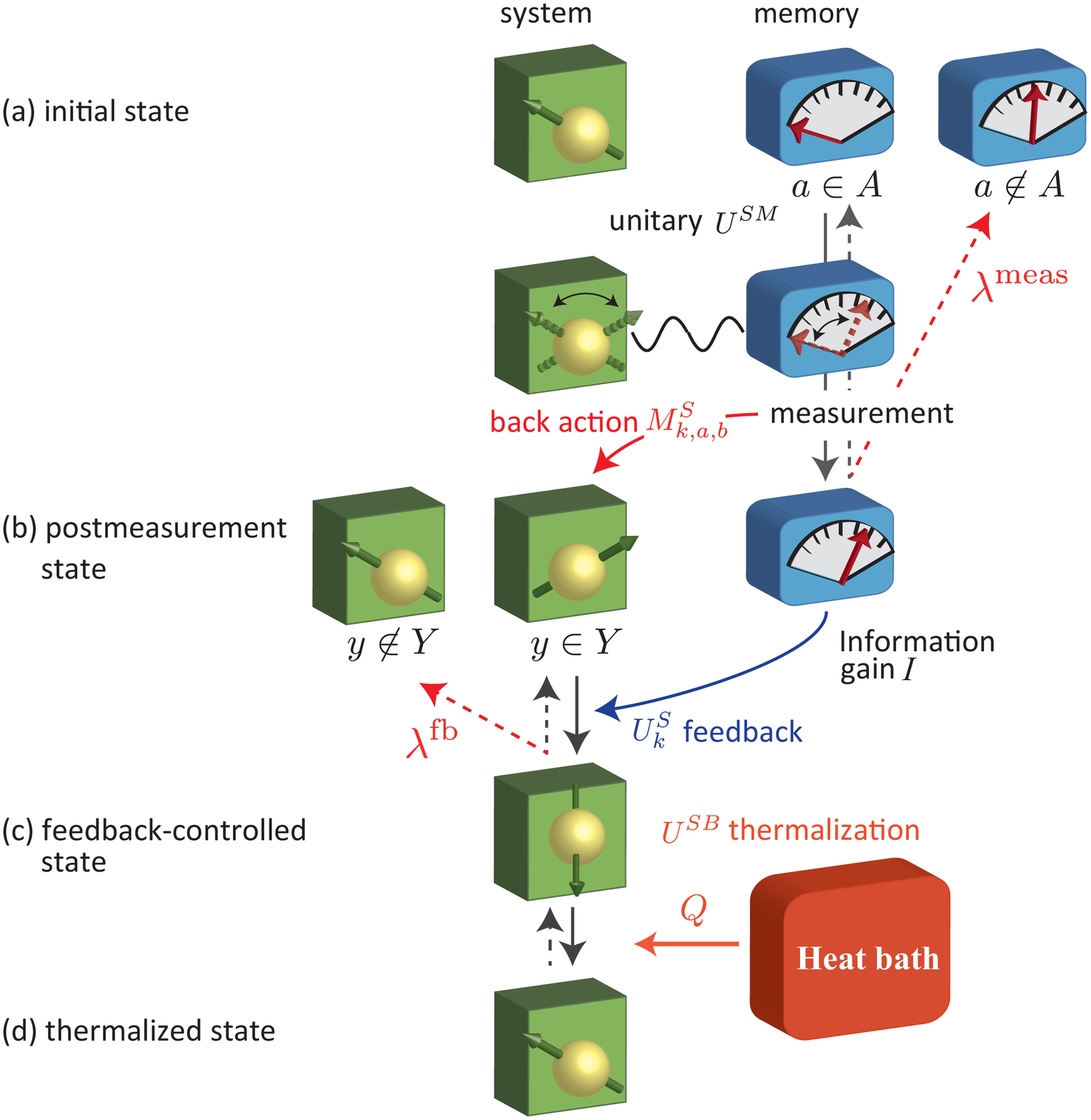}
\caption{\label{fig:qmsetup} Schematic illustration of the protocol. Solid downward arrows indicate the forward process and the dashed upward arrows show the backward process. {\it Forward process}: (a) We consider an initial state of the system described by the density matrix  $\rho^{S}_{\mathrm{ini}}=\sum_{x}p^{S}_{\mathrm{ini}}(x)\kket{\psi^{S}(x)}\bbra{\psi^{S}(x)}$ and that of the memory described by  $\rho^{M}_{\mathrm{ini}}=\sum_{a}p^{M}_{\mathrm{ini}}(a)\kket{\psi^{M}(a)}\bbra{\psi^{M}(a)}$. For the sake of simplicity of explanation, we show the case with $p^{S}_{\mathrm{ini}}(x)\neq 0$ for all $x$ and $p^{M}_{\mathrm{ini}}(a)\neq 0$ for only when $a\in A$ which is also the situation we consider in deriving quantum Jarzynski equalities. See the main text for a general case. Then, the support of $\rho^{M}_{\mathrm{ini}}$ belongs to the subspace ${\bf H}^{M}_{A}=\sum_{a\in A}\ket{\psi^{M}(a)}\bra{\psi^{M}(a)}$. (b) We first implement a general measurement $M^{S}_{k,a,b}$ by correlating $S$ and $M$ via $U^{SM}$, and then perform a projection $P^{M}_{k}=\sum_{b}\kket{\phi^{M}_{k}(b)}\bbra{\phi^{M}_{k}(b)}$ on $M$. From the measurement outcome $k$, we acquire the information about the system which is quantified by the information gain $\av{I}$. The support of the postmeasurement state $\rho^{S}(k)$ belongs to the subspace ${\bf H}^{S}_{k,Y}=\sum_{y\in Y}\kket{\varphi^{S}_{k}(y)}\bbra{\varphi^{S}_{k}(y)}$. (c) We perform a unitary transformation $U^{S}_{k}$, which realizes a feedback control to reduce the entropy of the system. (d) We attach a heat bath $\rho^{B}_{\mathrm{can}}$ to the system and let $S$ and $B$ interact via a unitary operator $U^{SB}$. During the protocol, heat is taken from $B$ and the amount of entropy reduction of $S$ via feedback is converted into work by a thermalization process. {\it Backward process}: (1) Backward process of the system (feedback control). We introduce the reference state $\rho^{S}_{\mathrm{r}}(k)\otimes\rho^{B}_{\mathrm{r}}$ as the initial state of the backward process. We reverse the protocol by applying the unitary operator $U^{\dagger S}_{k}U^{\dagger SB}$. After the protocol, the support of the density matrix of $S$ falls outside of the subspace ${\bf H}^{S}_{k,Y}$ in general. The total probability that the density matrix of the backward process ends up outside of the subspace ${\bf H}^{S}_{k,Y}$ (shown by the red dashed arrow) is denoted by $\lambda^{\mathrm{fb}}$ which measures the degree of absolute irreversibility of the feedback process. The process that returns to the subspace ${\bf H}^{S}_{k,Y}$ (shown by the gray dashed arrow) is an ordinary irreversible process, for which the ratio between the forward and backward probabilities can be related to the entropy production-like quantity and the information content as shown in Eq.~(\ref{sigmaplusi}). (2) Backward process of the memory. The initial state of the backward process is given by $\rho^{S}(k)\otimes\rho^{M}_{\mathrm{r}}(k)$. We implement the backward process by undoing the correlation via a unitary transformation $U^{\dagger SM}$, and the support of the density matrix of $M$ falls outside the subspace ${\bf H}^{M}_{A}$. The total probability that the density matrix of the backward process ends up outside the subspace ${\bf H}^{M}_{A}$ (shown by the red dashed arrow) is denoted by $\lambda^{\mathrm{meas}}$ which measures the degree of absolute irreversibility of the measurement process. The process that returns to the subspace ${\bf H}^{M}_{A}$ (shown by the gray dashed arrow) represents an ordinary irreversible process for which the ratio between the forward and backward probabilities can be related to the entropy production-like quantity and the information content given in Eq.~(\ref{sigmaminusi}).}
\end{center}
\end{figure}

To realize a general measurement and a feedback protocol, we consider the following protocol which is basically the same as the one considered in Ref.~\cite{Funo} and schematically illustrated in Fig.~\ref{fig:qmsetup}.

The total system consists of the system ($S$), the memory ($M$), the bath ($B$), and the interactions between them ($SM$ and $SB$). The corresponding Hamiltonian is given by
\beq
H=H_{k}^{S}(t)+V^{SM}(t)+ H^{M}+V^{SB}(t) +H^{B},
\eeq
where the interaction between the system and the heat bath is turned off until the thermalization process (e) starts. The  Hamiltonian of the system is controlled by the protocol that depends on the measurement outcome $k$ after the measurement step (b) at time $t=t_{\mathrm{meas}}$: 
\beq
H_{k}^{S}(t)=\begin{cases} H^{S}(t), &\hspace{5mm} 0\leq t\leq t_{\mathrm{meas}}; \\ H^{S}_{k}(t), &\hspace{5mm} t_{\mathrm{meas}}\leq t. \end{cases}
\eeq
We denote the initial Hamiltonian of the system by $H^{S}_{\mathrm{ini}}=H^{S}_{k}(t=0)$.

(a) Let the initial state of the system and the memory be
\beq
\rho^{SM}_{\mathrm{ini}}=\rho^{S}_{\mathrm{ini}}\otimes\rho^{M}_{\mathrm{ini}}.
\eeq

(b) A general quantum measurement on the system is implemented by performing a unitary transformation $U^{SM}=\text{T}\exp \left[-\ii\int^{t_{\mathrm{meas}}}_{0}(H^{S}_{k}(t)+V^{SM}(t)+H^{M}) \dd t \right]$ between the system and the memory followed by a projection $P^{M}_{k}=\sum_{b}\kket{\phi^{M}_{k}(b)}\bbra{\phi^{M}_{k}(b)}$ on the memory, where $\text{T}$ is the time-ordering operator. Here $P^{M}_{k}P^{M}_{l}=\delta_{k,l}P^{M}_{k}$ and $\{\kket{\phi^{M}_{k}(b)}\}_{k,b}$ is an orthonormal basis set of $M$. The postmeasurement state for the measurement outcome $k$ is given by
\beq
\rho^{SM}(k)=\frac{1}{p_{k}} P^{M}_{k} U^{SM}(\rho_{\mathrm{ini}}^{S}\otimes\rho^{M}_{\mathrm{ini}})U^{\dagger SM} P^{M}_{k} ,
\eeq
where $p_{k}=\Tr[P^{M}_{k} U^{SM}(\rho_{\mathrm{ini}}^{S}\otimes\rho^{M}_{\mathrm{ini}})U^{\dagger SM} P^{M}_{k}]$ is the probability of obtaining outcome $k$. The reduced density matrix of the system $\rho^{S}(k):=\Tr_{M}[\rho^{SM}(k)]$ is given by
\beq
\rho^{S}(k)=\sum_{a,b}\frac{M^{S}_{k,a,b}\rho^{S}M^{\dagger S}_{k,a,b}}{p_{k}}, \label{poststateabk}
\eeq
where 
\beq
M^{S}_{k,a,b}=\sqrt{p^{M}_{\mathrm{ini}}(a)}\bbra{\phi^{M}_{k}(b)}U^{SM}\kket{\psi^{M}(a)} \label{measoperator}
\eeq
is the measurement operator satisfying completeness relation 
\beq
\sum_{k,a,b}M^{\dagger S}_{k,a,b}M^{S}_{k,a,b}=1.
\eeq
Here $\kket{\psi^{M}(a)}$ and $p^{M}_{\mathrm{ini}}(a)$ in Eq.~(\ref{measoperator}) are given by the spectral decomposition of the initial state of the memory: $\rho^{M}_{\mathrm{ini}}=\sum_{a}p^{M}_{\mathrm{ini}}(a)\kket{\psi^{M}(a)}\bbra{\psi^{M}(a)}$. Note that for the special case of $\rho_{\mathrm{ini}}^{M}=\kket{\psi^{M}(0)}\bbra{\psi^{M}(0)}$ and $P^{M}_{k}=\kket{\psi^{M}(k)}\bbra{\psi^{M}(k)}$, the measurement is a pure measurement (which maps a pure state into a pure state)
\beq
M^{S}_{k}=\bbra{\psi^{M}(k)}U^{SM}\kket{\psi^{M}(0)}, \label{puremeasa}
\eeq
and the postmeasurement state is given by 
\beq
\rho^{S}(k)=p_{k}^{-1}M^{S}_{k}\rho^{S}_{\mathrm{ini}}M^{\dagger S}_{k}. \label{puremeasb}
\eeq

(c) We perform a unitary transformation $U^{S}_{k}$ depending on the measurement outcome $k$. Here the unitary operator is given by $U^{S}_{k}=\text{T}\exp[-\ii\int_{t_{\mathrm{meas}}}^{t_{\mathrm{fb}}}H^{S}_{k}(t)\dd t]$. We note that the above unitary operation associated with the measurement outcome is nothing but the feedback control. The density matrix of the system after the feedback control is given by
\beq
\rho^{S}_{\mathrm{fb}}(k)=U^{S}_{k}\rho^{S}(k) U^{\dagger S}_{k}.
\eeq

(d) Finally, we let the system and heat bath interact with each other so that the reduced entropy of the system via feedback control is converted to heat. Here, we assume that the initial state of the heat bath is given by the canonical distribution, i.e., $\rho^{B}_{\mathrm{can}}=\exp[-\beta(H^{B}-F^{B})]$. The final state is given by
\beq
\rho^{SB}_{\mathrm{fin}}(k)=U^{SB}_{k}(\rho^{S}_{\mathrm{fb}}(k)\otimes\rho^{B}_{\mathrm{can}})U^{\dagger SB}_{k},
\eeq
where the interaction between $S$ and $B$ discribed by the unitary operator $U^{SB}_{k}=\text{T}\exp[-\ii \int_{t_\mathrm{fb}}^{t_{\mathrm{fin}}}( H_{k}^{S}(t)+V^{SB}(t) +H^{B})\dd t]$, which may, in general, depend on $k$.

Now we introduce reference states for each subsystem and define entropy production-like quantities which measure the amount of entropy of $SB$ ($M$) that is reduced (or produced) due to the feedback control (or measurement). The reference states of each subsystem is given by 
\beqa
\rho^{S}_{\mathrm{r}}(k)&=&\sum_{z}p^{S}_{\mathrm{r}}(z|k)\kket{\phi^{S}_{k}(z)}\bbra{\phi^{S}_{k}(z)},\label{specrefS}\\
\rho^{M}_{\mathrm{r}}(k)&=&\sum_{b}p^{M}_{\mathrm{r}}(b|k)\kket{\phi^{M}_{k}(b)}\bbra{\phi^{M}_{k}(b)}, \label{specrefM}\\
\rho^{B}_{\mathrm{r}}&=&\sum_{i}p^{B}_{\mathrm{can}}(j)\kket{\psi^{B}(j)}\bbra{\psi^{B}(j)},\label{specrefB}
\eeqa
where $p^{B}_{\text{can}}(j)=e^{-\beta(E^{B}(j)-F^{B})}$ is the canonical distribution, and $E^{B}(j)$ and $\kket{\psi^{B}(j)}$ is the eigenenergy and energy eigenstate of the heat bath, respectively.

We define the following quantity that measures the amount of entropy reduction of $SB$ due to feedback control:
\beqa
\left\langle \sigma^{SB}\right\rangle &=& -S(\rho^{S}_{\mathrm{ini}}\otimes\rho^{B}_{\mathrm{can}})-\sum_{k}p_{k}\Tr[\rho^{SB}_{\mathrm{fin}}(k)\ln(\rho^{S}_{\mathrm{r}}(k)\otimes\rho^{B}_{\mathrm{r}})]\\
&=&-S(\rho^{S}_{\mathrm{ini}})-\sum_{k}p_{k}\Tr[\rho^{S}_{\mathrm{fin}}(k)\ln\rho^{S}_{\mathrm{r}}(k)]-\beta \left\langle Q\right\rangle, \label{entproSB}
\eeqa
where
\beq
\left\langle Q\right\rangle =\Tr[H^{B}\rho^{B}_{\mathrm{can}}]-\sum_{k}p_{k}\Tr[H^{B}\rho^{B}_{\mathrm{fin}}(k)]
\eeq
is the energy change of the heat bath which we identify as heat transfered from $B$ to $S$. Note that if we choose the reference state as the final density matrix of $S$, i.e., $\rho^{S}_{\mathrm{r}}(k)=\rho^{S}_{\mathrm{fin}}(k)$, Eq.~(\ref{entproSB}) is nothing but the total entropy change of $SB$ due to feedback control:
\beq
\left\langle\sigma^{SB}\right\rangle=\Delta S^{S}-\beta\left\langle Q\right\rangle,
\eeq
where 
\beq
\Delta S^{S}=\sum_{k}p_{k}S(\rho^{S}_{\mathrm{fin}}(k))-S(\rho^{S}_{\mathrm{ini}})
\eeq
is a change in the von Neumann entropy of the system during the entire protocol. 

We also define the following quantity which measures the amount of entropy produced in $M$ due to measurement:
\beq
\left\langle\sigma^{M}\right\rangle=-S(\rho^{M}_{\mathrm{ini}})-\Tr[\rho^{M}_{\mathrm{fin}}\ln\rho^{M}_{\mathrm{r}}], \label{entproM}
\eeq
where $\rho^{M}_{\mathrm{fin}}=\sum_{k}p_{k}\rho^{M}(k)$ is the final density matrix of $M$ and $\rho^{M}_{\mathrm{r}}:=\sum_{k}p_{k}\rho_{\mathrm{r}}^{M}(k)$. If we choose the reference state as the canonical distribution, Eqs.~(\ref{entproSB}) and (\ref{entproM}) are related to work and the free-energy difference, respectively, as shown in the next section.

The entropy production-like quantities~(\ref{entproSB}) and (\ref{entproM}) contain not only the effect of dissipated entropy due to irreversibility of the process but also the effect of entropy change due to information processing (measurement and feedback control), and they can take either positive or negative values depending on the process. The effect of the information exchange between the system and the memory can be expressed by the information gain (quantum-classical mutual information) of the system $S$~\cite{Groenewold,Ozawa,Sagawa1}:
\beq
\av{I}=S(\rho_{\mathrm{ini}}^{S})-\sum_{k}p_{k}S(\rho^{S}(k)), \label{infogain}
\eeq
which is the amount of entropy that is reduced from the system due to the measurement. The information gain is bounded from above by the Shannon entropy $H=-\sum_{k}p_{k}\ln p_{k}$, i.e., $\av{I}\leq H$, where the equality holds if and only if the measurement is given by a projective measurement using the diagonal basis of $\rho^{S}_{\mathrm{ini}}$. Moreover, the information gain is nonnegative for any premeasurement state if the measurement is a pure measurement~(\ref{puremeasa}) as discussed in Ref.~\cite{Ozawa}.

Extracting the information gain from entropy production-like quantities~(\ref{entproSB}) and (\ref{entproM}), we obtain the measures of irreversibility during measurement and feedback processes. For the feedback process, we have
\beqa
\aav{\sigma^{SB}}+\av{I}&=& -S(\rho^{B}_{\mathrm{can}})-\sum_{k}p_{k}S(\rho^{S}(k))-\sum_{k}p_{k}\Tr[\rho^{SB}_{\mathrm{fin}}(k)\ln(\rho^{S}_{\mathrm{r}}(k)\otimes\rho^{B}_{\mathrm{r}})] \nonumber \\
&=&-\sum_{k}p_{k}\left\{S(\rho^{S}(k)\otimes\rho^{B}_{\mathrm{can}})+\Tr[(\rho^{S}(k)\otimes\rho^{B}_{\mathrm{can}})\ln\tilde{\rho}^{SB}_{\mathrm{r}}(k)]\right\} \nonumber \\
&=& \sum_{k}p_{k}S(\rho^{S}(k)\otimes \rho^{B}_{\mathrm{can}}||\tilde{\rho}^{SB}_{\mathrm{r}}(k))\geq 0, \label{entproinfineq}
\eeqa
where $\tilde{\rho}_{\mathrm{r}}^{SB}(k)=U^{\dagger S}_{k}U^{\dagger SB}_{k} (\rho^{S}_{\mathrm{r}}(k)\otimes \rho^{B}_{\mathrm{r}})U^{SB}_{k}U^{S}_{k}$ is the final density matrix of the backward process by reversing the thermalization and feedback control protocols. Note that the feedback protocol of the system (and the heat bath) is reversible if and only if $\rho^{S}(k)\otimes \rho^{B}_{\mathrm{can}}=\tilde{\rho}_{\mathrm{r}}^{SB}(k)$~\cite{feedrev}, which is the equality condition of the last inequality~(\ref{entproinfineq}), that is $\aav{\sigma^{SB}}+\av{I}=0$. 

Similarly, for the measurement process, we have
\beqa
\aav{\sigma^{M}}-\av{I}&=&-S(\rho^{SM}_{\mathrm{ini}})-\Tr[\rho^{M}_{\mathrm{fin}}\ln\rho^{M}_{\mathrm{r}}]+\sum_{k}p_{k}S(\rho^{S}(k)) \nonumber \\
&=&-S(U^{SM}\rho^{SM}_{\mathrm{ini}}U^{\dagger SM})-\Tr\left[\rho_{\text{meas}}^{SM}\ln\left(\sum_{k}p_{k}\rho^{S}(k)\otimes\rho^{M}_{\mathrm{r}}(k)\right)\right] \nonumber \\
&=&\Delta S_{\text{meas}}^{SM}+S(\rho^{SM}_{\text{meas}}||\sum_{k}p_{k}\rho^{S}(k)\otimes\rho^{M}_{\mathrm{r}}(k))\geq 0, \label{sigmaipro}
\eeqa
where 
\beq
\rho^{SM}_{\text{meas}}=\sum_{k}p_{k}\rho^{SM}(k)
\eeq
is the average postmeasurement state over measurement outcomes, and
\beq
\Delta S^{SM}_{\mathrm{meas}}=S(\rho^{SM}_{\text{meas}})-S(U^{SM}\rho^{SM}_{\mathrm{ini}}U^{\dagger SM}) \geq 0
\eeq
is a change in the von Neumann entropy due to projection $P^{M}_{k}$ and the inequality results from the fact that von Neumann entropy does not decrease under projection measurements. The nonnegativity in Eq.~(\ref{sigmaipro}) shows the irreversibility of the measurement process.

Next, let us consider the following spectral decompositions of the initial states of the system, the heat bath, and the memory 
\beqa
\rho^{S}_{\mathrm{ini}}&=&\sum_{x}p^{S}_{\mathrm{ini}}(x)\kket{\psi^{S}(x)}\bbra{\psi^{S}(x)}, \\ 
\rho^{B}_{\mathrm{ini}}&=&\sum_{i}p^{B}_{\mathrm{can}}(h)\kket{\psi^{B}(h)}\bbra{\psi^{B}(h)}, \\
\rho_{\mathrm{ini}}^{M}&=&\sum_{a}p^{M}_{\mathrm{ini}}(a)\kket{\psi^{M}(a)}\bbra{\psi^{M}(a)}. 
\eeqa
Let us also decompose the postmeasurement state of the system as follows: 
\beq
\rho^{S}(k)=\sum_{y}p^{S}(y|k)\kket{\varphi^{S}_{k}(y)}\bbra{\varphi^{S}_{k}(y)}. \label{diagpost}
\eeq
Using the above decompositions, we calculate Eqs.~(\ref{entproinfineq}) and (\ref{sigmaipro}) and define an unaveraged form of Eqs.~(\ref{entproSB}), (\ref{entproM}) and (\ref{infogain}), along a line similar to what we did in deriving the unaverage form of the entropy production in Eq.~(\ref{sigmadec}). From Eq.~(\ref{entproinfineq}), we obtain
\beqa 
\aav{\sigma^{SB}}&=&\sum_{x}p^{S}_{\mathrm{ini}}(x)\ln p^{S}_{\mathrm{ini}}(x)+\sum_{h}p^{B}_{\mathrm{can}}(h)\ln p^{B}_{\mathrm{can}}(h) \nonumber \\
& &-\sum_{k,j,z}p_{k} \bbra{\phi^{S}_{k}(z)}\otimes\bbra{\psi^{B}(j)}\rho^{SB}_{\mathrm{fin}}(k) \kket{\phi^{S}_{k}(z)}\otimes\kket{\psi^{B}(j)}\ln (p^{S}_{\mathrm{r}}(z|k)p^{B}_{\mathrm{can}}(j) )      \nonumber \\
&=&\sum_{x,a,h,k,y,b,j,z}p(x,a,h,k,y,b,j,z)\ln\frac{p^{S}_{\mathrm{ini}}(x)p^{B}_{\mathrm{can}}(h)}{p^{S}_{\mathrm{r}}(z|k)p^{B}_{\mathrm{can}}(j)}, \label{systemdec}
\eeqa
where we introduce the forward probability distribution corresponding to the forward process of the total system:
\beq
p(x,a,h,k,y,b,j,z)=p^{S}_{\mathrm{ini}}(x)p^{M}_{\mathrm{ini}}(a)p^{B}_{\mathrm{can}}(h)p(k,y,b|x,a)p(z,j|k,y,h), \label{totforwardprob}
\eeq
where
\beqa
p(k,y,b|x,a)&=&|  \bbra{ \varphi^{S}_{k}(y)}\otimes\bbra{\phi^{M}_{k}(b)} U^{SM} \kket{ \psi^{S}(x) }\otimes\kket{\psi^{M}(a)} |^{2} \nonumber \\
&=&\frac{1}{p^{M}_{\mathrm{ini}}(a)}|  \bbra{ \varphi^{S}_{k}(y)}M^{S}_{k,a,b}\kket{ \psi^{S}(x) } |^{2}
\eeqa
is the transition probability between the state labeled by $x,a$ to the state labeled by $k,y,b$ during the measurement process, and
\beq
p(z,j|k,y,h)=|\bbra{\phi^{S}_{k}(z)}\otimes\bbra{\psi^{B}(j)}U^{SB}_{k}U^{S}_{k}\kket{\varphi^{S}_{k}(y)}\otimes\kket{\psi^{B}(h)}|^{2}
\eeq
is the transition probability between the state labeled by $k,y,h$ to the state labeled by $z,j$ during the feedback and the thermalization protocol. Note that in deriving the last line of Eq.~(\ref{systemdec}), we used the relation
\beqa
& &\bbra{\phi^{S}_{k}(z)}\otimes\bbra{\psi^{B}(j)}\rho^{SB}_{\mathrm{fin}}(k) \kket{\phi^{S}_{k}(z)}\otimes\kket{\psi^{B}(j)} \nonumber \\
&=&\sum_{h,y}p(z,j|k,y,h)p^{S}(y|k)p^{B}_{\mathrm{can}}(h) \nonumber \\
&=&\sum_{x,a,h,y,b}p_{k}^{-1}p^{S}_{\mathrm{ini}}(x)p^{M}_{\mathrm{ini}}(a)p^{B}_{\mathrm{can}}(h)p(k,y,b|x,a)p(z,j|k,y,h).
\eeqa
We also follow the same procedure for the memory using~Eq.~(\ref{entproM}) and obtain
\beqa 
\aav{\sigma^{M}}&=&\sum_{a}p^{M}_{\mathrm{ini}}(a)\ln p^{M}_{\mathrm{ini}}(a)-\sum_{k,b}p_{k} \bbra{\phi^{M}_{k}(b)}\rho^{M}(k)\kket{\phi^{M}_{k}(b)} \ln (p_{k}p^{M}_{\mathrm{r}}(b|k))      \nonumber \\
&=&\sum_{x,a,k,y,b}p^{\mathrm{meas}}(x,a,k,y,b)\ln\frac{p^{M}_{\mathrm{ini}}(a)}{p_{k}p^{M}_{\mathrm{r}}(b|k)}, \label{memorydec}
\eeqa
where the forward probability of the measurement process is defined as
\beq
p^{\text{meas}}(x,a,k,y,b)=p^{S}_{\mathrm{ini}}(x)p^{M}_{\mathrm{ini}}(a)p(k,y,b|x,a).
\eeq

From Eqs.~(\ref{infogain}) (\ref{systemdec}) and (\ref{memorydec}), we define unaveraged entropy production-like quantities and the corresponding information content as follows: 
\beqa
\sigma^{SB}(x,h,k,j,z)&=& \ln [p^{S}_{\mathrm{ini}}(x)p^{B}_{\mathrm{can}}(h)] -\ln [p^{S}_{\mathrm{r}}(z|k)p^{B}_{\mathrm{can}}(j)] \\
&=&\ln p^{S}_{\mathrm{ini}}(x)-\ln p^{S}_{\mathrm{r}}(z|k) -\beta Q(h,j) ,  \label{EPS} \\
\sigma^{M}(a,k,b)&=&\ln p^{M}_{\mathrm{ini}}(a)-\ln [p_{k}p^{M}_{\mathrm{r}}(b|k)], \label{EPM} \\
I(x,k,y)&=&\ln p^{S}(y|k)- \ln p^{S}_{\mathrm{ini}}(x) ,  \label{genejarqcmutualdef}
\eeqa
where $Q(h,j)=E^{B}(h)-E^{B}(j)$ is the heat transfered from the heat bath to the system. Since the entropy production relates the forward and reference probabilities as in Eq.~(\ref{entratio}), we have similar relations for the combinations $\sigma^{SB}+I$ and $\sigma^{M}-I$. First by decomposing Eq.~(\ref{entproinfineq}), we have
\beq
\aav{\sigma^{SB}}+\aav{I}=\sum_{h,k,y,j,z}p^{\mathrm{fb}}(h,k,y,j,z)\ln\frac{p^{S}(y|k)p^{B}_{\mathrm{can}}(h)}{p^{S}_{\mathrm{r}}(z|k)p^{B}_{\mathrm{can}}(j)},
\eeq
 where
 \beq
p^{\mathrm{fb}}(h,k,y,j,z)= p^{B}_{\mathrm{can}}(h)p^{S}(y|k)p_{k}p(z,j|k,y,h) \label{forwardfb}
\eeq
is the forward probability of the feedback control process. Then, $\sigma^{SB}+I$ and $\sigma^{M}-I$ are expressed in terms of the logarithm of the ratio between the forward and reference probabilities as
\beqa
\sigma^{SB}(x,h,k,j,z)+I(x,k,y)&=&\ln\frac{p^{\mathrm{fb}}(h,k,y,j,z)}{p^{\mathrm{fb}}_{\mathrm{r}}(h,k,y,j,z)}, \label{sigmaplusi}\\
\sigma^{M}(a,k,b)-I(x,k,y)&=&\ln\frac{p^{\mathrm{meas}}(x,a,k,y,b)}{p^{\mathrm{meas}}_{\mathrm{r}}(x,a,k,y,b)}, \label{sigmaminusi}
\eeqa
where
\beq
p^{\mathrm{fb}}_{\mathrm{r}}(h,k,y,j,z)=p_{k}p^{S}_{\mathrm{r}}(z|k)p^{B}_{\mathrm{can}}(j)p(z,j|k,y,h) \label{fbrefprob} 
\eeq
and
\beq
p^{\text{meas}}_{\mathrm{r}}(x,a,k,y,b)=p_{k}p^{S}(y|k)p^{M}_{\mathrm{r}}(b|k)p(k,y,b|x,a) \label{refmeasprob}
\eeq
are the reference probabilites of the feedback and measurement processes, respectively. 

 Note that Eq.~(\ref{fbrefprob}) gives the probability of the system and the heat bath returning to the postmeasurement state of the system and the initial state of the heat bath $\kket{\varphi^{S}_{k}(y)}\otimes\kket{\psi^{B}(h)}$ when we start from the initial state of the backward process $\rho^{S}_{\mathrm{r}}(z|k)\otimes\rho^{B}_{\mathrm{can}}$ and do the reverse of the thermalization and feedback control $U^{\dagger S}_{k}U^{\dagger SB}_{k}$, as shown in the gray dashed upward arrow in Fig.~\ref{fig:qmsetup}. Also, Eq.~(\ref{refmeasprob}) gives the probability of the system and the memory returning to the initial state $\kket{\psi^{S}(x)}\otimes\kket{\psi^{M}(a)}$ when we start from the initial state of the backward process $\sum_{k}p_{k}\rho^{S}(k)\otimes\rho^{M}_{\mathrm{r}}(k)$ and let the system and the memory undo the correlation by applying a unitary operation $U^{\dagger SM}$, as shown by the gray dashed upward arrow in Fig.~\ref{fig:qmsetup}. We use the definitions of entropy production-like quantities~(\ref{EPS}) and (\ref{EPM}) and the information content~(\ref{genejarqcmutualdef}) to derive quantum fluctuation theorems for both the feedback-controlled system and the measurement device.
%
\subsection{Quantum fluctuation theorems with feedback control}

We derive quantum fluctuation theorems for both the feedback-controlled system and the measurement device from the fact that the sum of the reference probabilities is unity for both the feedback control process~(\ref{fbrefprob}) and the measurement process~(\ref{refmeasprob}):
\beqa
1&=&\sum_{h,k,y,j,z}p^{\mathrm{fb}}_{\mathrm{r}}(h,k,y,j,z), \label{sumreffb}\\
1&=&\sum_{x,a,y,k,b}p^{\text{meas}}_{\mathrm{r}}(x,a,k,y,b). \label{sumrefmeas}
\eeqa
As in Eq.~(\ref{refprobdec}), we decompose Eqs.~(\ref{sumreffb}) and (\ref{sumrefmeas}) into two parts; one is the part where we can take the ratio between the forward and reference probabilities, and the other is the part where the corresponding forward probability vanishes. (See dashed upward arrows in Fig.~\ref{fig:qmsetup}. )

We introduce a set of labels corresponding to the non-vanishing probability distributions as follows: we introduce $Y$ as a set of labels $y$ satisfying $p(y|k)\neq 0$, and $A$ as a set of labels $(x,a)$ satisfying both $p^{S}_{\mathrm{ini}}(x)\neq 0$ and $p^{M}_{0}(a)\neq 0$. Then, the support of the postmeasurement state of the system belongs to the subspace ${\bf H}^{S}_{k,Y}$, which is spanned by $\{\kket{\varphi^{S}_{k}(y)}\}_{y\in Y}$, and the support of the initial state of the system and the memory belongs to the subspace ${\bf H}^{SM}_{A}$, which is spanned by $\{\kket{\psi^{S}(x)}\otimes\kket{\psi^{M}(a)}\}_{(x,a)\in A}$.

Using the above notations, we decompose the reference states and derive quantum fluctuation theorems as follows: for the system and the heat bath, we have
\beqa
1&=&\sum_{h,k,y\not\in Y,j,z}p^{\mathrm{fb}}_{\mathrm{r}}(h,k,y,j,z)+\sum_{h,k,y\in Y,j,z}p^{\mathrm{fb}}_{\mathrm{r}}(h,k,y,j,z)\nonumber \\
&=&\lambda^{\mathrm{fb}}+\sum_{h,k,y\in Y,j,z}p^{\mathrm{fb}}(h,k,y,j,z)\frac{p^{\mathrm{fb}}_{\mathrm{r}}(h,k,y,j,z)}{p^{\mathrm{fb}}(h,k,y,j,z)} \nonumber \\
&=&\lambda^{\mathrm{fb}}+\sum_{x,a,h,k,y\in Y,b,j,z}p(x,a,h,k,y,b,j,z)\ee^{-\sigma^{SB}(x,h,k,j,z)-I(x,k,y)}\nonumber \\
&=&\lambda^{\mathrm{fb}}+\aav{\ee^{-\sigma^{SB}-I}},\label{proofents}
\eeqa
where we used Eq.~(\ref{sigmaplusi}) in deriving the third equality. Here 
\beq
\lambda^{\mathrm{fb}}=\sum_{h,k,y\not\in Y,j,z}p^{\mathrm{fb}}_{\mathrm{r}}(h,k,y,j,z) \label{lambdas}
\eeq
is the sum of the reference probabilities, where the density matrix of the backward process ends up outside of the subspace ${\bf H}^{S}_{k,Y}$, and the overlap with the postmeasurement state of the forward process is zero.

For the memory, the decomposition of the reference probability leads to
\beqa
1&=&\sum_{(x,a)\not\in A,k,y,b}p^{\text{meas}}_{\mathrm{r}}(x,a,k,y,b)+\sum_{(x,a)\in A,k,y,b} p^{\text{meas}}_{\mathrm{r}}(x,a,k,y,b) \nonumber \\
&=&\lambda^{\mathrm{meas}}+\sum_{(x,a)\in A,h,k,y,b,j,z}p(x,a,h,k,y,b,j,z)\frac{p^{\text{meas}}_{\mathrm{r}}(x,a,k,y,b)}{p^{\text{meas}}(x,a,k,y,b)}\nonumber \\
&=&\lambda^{\mathrm{meas}}+\sum_{(x,a)\in A,h,k,y,b,j,z}p(x,a,h,k,y,b,j,z)\ee^{-\sigma^{M}(a,k,b)+I(x,k,y)} \nonumber \\
&=&\lambda^{\mathrm{meas}}+\aav{\ee^{-\sigma^{M}+I}},\label{proofentm}
\eeqa
where we used Eq.~(\ref{sigmaminusi}) in deriving the third equality. Here
\beq
\lambda^{\mathrm{meas}}=\sum_{(x,a)\not\in A,k,y,b}p^{\text{meas}}_{\mathrm{r}}(x,a,k,y,b)
\eeq
is the sum of the reference probabilities such that the density matrix of the backward process ends up outside of the subspace ${\bf H}^{SM}_{A}$ and the overlap with the initial state of the forward process vanishes.

Rewriting Eqs.~(\ref{proofents}) and (\ref{proofentm}), we obtain the quantum fluctuation theorems for the system and the memory:
\beq
\aav{\ee^{-\sigma^{SB}-I}}=1-\lambda^{\mathrm{fb}}, \label{QCfluc}
\eeq
\beq
\aav{\ee^{-\sigma^{M}+I}}=1-\lambda^{\mathrm{meas}}.  \label{QMfluc}
\eeq
Using the Jensen's inequality, we can reproduce second law-like inequalities by using Eqs.~(\ref{QCfluc}) and (\ref{QMfluc}):
\beq
\aav{\sigma^{SB}}\geq -\av{I}-\ln(1-\lambda^{\mathrm{fb}}), \label{SBineqfb}
\eeq
\beq
\aav{\sigma^{M}}\geq \av{I}-\ln(1-\lambda^{\mathrm{meas}}), \label{Mineqmeas}
\eeq
where the presence of absolute irreversibility (nonzero $\lambda$) imposes stronger lower bounds on $\aav{\sigma^{SB}}$ and $\aav{\sigma^{M}}$ compared with the previous results~(\ref{entproinfineq}) and (\ref{sigmaipro}) given in Ref.~\cite{Sagawa5}. Since $\lambda^{\mathrm{fb}}$ gives the total probability of the density matrix of the backward process ending up outside of the subspace ${\bf H}^{S}_{k,Y}$, it measures the degree of absolute irreversibility of the feedback protocol. 

If the measurement on the system is given by projective measurements $\ket{k}\bra{k}_{S}$, the situation becomes simple. In this case,  $\lambda^{\mathrm{fb}}$ takes the following form:
\beqa
\lambda^{\mathrm{fb}}&=&\sum_{h,k\neq y,j,z}p_{k}p^{S}_{\mathrm{r}}(z|k)p_{\mathrm{r}}(j)|\bbra{y}_{S}\otimes\bbra{\psi^{B}(h)}U^{\dagger S}_{k}U^{\dagger SB}_{k}\kket{\phi^{S}_{k}(z)}\otimes\kket{\psi^{B}(j)}|^{2} \nonumber \\
&=&\sum_{k\neq y}p_{k} \left\langle y \left | \Tr_{B}[U^{\dagger S}_{k}U^{\dagger SB}_{k}(\rho^{S}_{\mathrm{r}}(k)\otimes\rho^{B}_{\mathrm{can}})U^{SB}_{k}U^{S}_{k} ] \right| y\right\rangle_{S}, \label{projectionlambdafb}
\eeqa
which is the sum of the probability of the backward protocol for each measurement outcome $k$ that does not end in the state $\ket{k}_{S}$. If the unitary operator brings the postmeasurement state $\ket{k}_{S}$ into the reference state $\rho^{S}(k)$ for all $k$, the feedback (and thermalization) process is reversible and $\lambda^{\mathrm{fb}}$ vanishes; otherwise the irreversibility of the process reduces the efficiency of the feedback gain. Note that Eq.~(\ref{QCfluc}) holds even for projective measurements on the system, where the previous results in Ref.~\cite{Funo} are inapplicable, since we take into account the effect of absolute irreversibility. Although the obtained information is given by the Shannon entropy and is maximal for projective measurements, feedback protocol tends to be absolutely irreversible since the postmeasurement state is sharply localized in the Hilbert space; it is given by a pure state $\ket{k}$. Similarly, $\lambda^{\mathrm{meas}}$ measures the absolute irreversibility of the measurement process since it is nonzero when the density matrix of the backward protocol ends up outside of the subspace ${\bf H}^{SM}_{A}$.

Now let us compare the obtained equalities~(\ref{QCfluc}) and (\ref{QMfluc}) with the quantum fluctuation theorems of a total system by using the total entropy production $\sigma_{\mathrm{tot}}(x,h,k,j,a,b)=\sigma^{SB}(x,h,k,j,a)+\sigma^{M}(a,k,b)$. The total entropy production can be written in the form
\beqa
\sigma_{\mathrm{tot}}(x,h,k,j,a,b)&=&\ln\frac{p^{S}_{\mathrm{ini}}(x)p^{B}_{\mathrm{can}}(h)p^{M}_{\mathrm{ini}}(a)}{p^{S}_{\mathrm{r}}(z|k)p^{B}_{\mathrm{can}}(j)p_{k}p^{M}_{\mathrm{r}}(b|k)}\nonumber \\
&=&\ln\frac{p(x,a,h,k,b,j,z)}{p_{\mathrm{r}}(x,a,h,k,b,j,z)},
\eeqa
where we used the total probability distribution of the forward process~(\ref{totforwardprob}), and
\beq
p_{\mathrm{r}}(x,a,h,k,b,j,z)=p^{S}_{\mathrm{r}}(z|k)p^{B}_{\mathrm{can}}(j)p_{k}p^{M}_{\mathrm{r}}(b|k)p(z,j|k,y,h)p(k,y,b|x,a) \label{totrefprob}
\eeq
is the total probability distribution of the backward process. Since the sum of the reference probability~(\ref{totrefprob}) is unity, we can derive the quantum fluctuation theorem for the total system:
\beqa
1&=&\sum_{(x,a)\not\in A,h,k,b,j,z}p_{\mathrm{r}}(x,a,h,k,b,j,z)+\sum_{(x,a)t\in A,h,k,b,j,z}p_{\mathrm{r}}(x,a,h,k,b,j,z) \nonumber \\
&=&\lambda^{\mathrm{tot}}+\sum_{(x,a)\in A,h,k,b,j,z}p(x,a,h,k,b,j,z)\ln\frac{p_{\mathrm{r}}(x,a,h,k,b,j,z)}{p(x,a,h,k,b,j,z)} \nonumber \\
&=&\lambda^{\mathrm{tot}}+\aav{\ee^{-\sigma_{\mathrm{tot}}}}. \label{totflucthm}
\eeqa
Since the obtained fluctuation theorem is applicable to the total system, the effect of information exchange between $S$ and $M$ is canceled, and the information content does not appear in Eq.~(\ref{totflucthm}). Moreover, $\lambda^{\mathrm{tot}}$ measures absolute irreversibility of the combined process of the measurement and feedback control, whereas from Eqs.~(\ref{QCfluc}) and (\ref{QMfluc}) we can separately obtain the information about the absolute irreversibility in measurement and feedback.

\subsection{\label{sec:QJE}Quantum Jarzynski equalities with feedback control}

In this subsection, we derive the quantum Jarzynski equality for the feedback-controlled system by assuming that the initial and reference states are given by canonical distributions
\beq
\rho^{S}_{\mathrm{ini}}=\ee^{-\beta (H^{S}_{\mathrm{ini}}-F^{S}_{\mathrm{ini}})}, \label{Jarassuma}
\eeq
and 
\beq
\rho^{S}_{\mathrm{r}}(k)=\ee^{-\beta(H^{S}_{\mathrm{fin}}(k)-F^{S}_{k})}, \label{Jarassumb}
\eeq
respectively, where $H^{S}_{\mathrm{ini}}$ and $H^{S}_{\mathrm{fin}}(k)$ are the initial and final Hamiltonians of the system. Then, the orthogonal bases $\{\kket{\psi^{S}(x)}\}$ and $\{\kket{\phi^{S}_{k}(z)}\}$ are given by the set of energy eigenfunctions: $H^{S}_{\mathrm{ini}}\kket{\psi^{S}(x)}=E^{S}_{\mathrm{ini}}(x)\kket{\psi^{S}(x)}$ and $H^{S}_{\mathrm{fin}}(k)\kket{\phi^{S}_{k}(z)}=E^{S}_{\mathrm{fin},k}(z)\kket{\phi^{S}_{k}(z)}$. Now $\sigma^{SB}$ is related to the work done by the system as follows:
\beqa
\sigma^{SB}(x,h,k,j,z)= -\beta\left[W^{S}(x,h,k,j,z)+\Delta f^{S}(k)\right], \label{entropywork}
\eeqa
where 
\beq
W^{S}(x,k,z)=E^{S}_{\mathrm{ini}}(x)-E_{\mathrm{fin},k}^{S}(z)+Q(h,j)
\eeq
is the work done by the system, and $\Delta f^{S}(k)=F^{S}_{k}-F^{S}_{\mathrm{ini}}$ is the free-energy difference.

We now derive the following quantum Jarzynski equality for a feedback-controlled system by using Eq.~(\ref{QCfluc}):
\beq
\aav{\ee^{\beta(W^{S}+\Delta f^{S})-I}}=1-\lambda^{\mathrm{fb}}. \label{QCGeneJar}
\eeq
Using the Jensen's inequality, Eq.~(\ref{QCGeneJar}) reproduces the generalized second law under feedback control:
\beq
\aav{W^{S}}\leq -\aav{\Delta f^{S}} +\bz T \av{I}+\bz T\ln(1-\lambda^{\mathrm{fb}}),\label{genejarseclaw}
\eeq
where 
\beq
\aav{W^{S}}= \Tr[\rho^{S}_{\mathrm{ini}}H^{S}_{\mathrm{ini}}]-\sum_{k}p_{k}\Tr[\rho^{S}_{\mathrm{fin}}(k)H^{S}_{\mathrm{fin}}(k)]+\left\langle Q\right\rangle \label{workfeedback}
\eeq
is the averaged work done by the system. Imperfect feedback control leads to nonzero $\lambda^{\mathrm{fb}}$, which lowers the extractable work from the system as shown in Eq.~(\ref{genejarseclaw}).

Next, we derive the quantum Jarzynski equality for the memory that acquires the measurement results by assuming that the initial and reference states are given by the canonical distributions
\beq
\rho^{M}_{\mathrm{ini}}=\ee^{-\beta(H^{M}_{0}-F^{M}_{0})}, \label{initialcanom}
\eeq
\beq
\rho^{M}_{\mathrm{r}}(k)=\ee^{-\beta(H^{M}_{k}-F^{M}_{k})}.
\eeq
We assume that the initial state of the memory~(\ref{initialcanom}) is given by the local equilibrium state defined as the canonical distribution using a local Hamiltonian $H^{M}_{0}$, where the Hamiltonian of the memory is decomposed into $H^{M}=\oplus_{k}H^{M}_{k}$~\cite{Sagawa2}. Here, the spectral decomposition of each local Hamiltonian is given by $H^{M}_{k}=\sum_{b}E^{M}_{k}(b)\kket{\phi^{M}_{k}(b)}\bbra{\phi^{M}_{k}(b)}$. For convenience, let us relabel $a$ as $a=(a_{1},a_{2})$ so that $\kket{\psi^{M}(a)}=\kket{\phi^{M}_{a_{1}}(a_{2})}$. Then $p^{M}_{\mathrm{ini}}(a)\neq 0$ if $a=(0,a_{2})$ and zero otherwise for the initial state defined in Eq.~(\ref{initialcanom}). Now $\sigma^{M}$ is related to the work done on the memory as follows:
\beq
\sigma^{M}(a,k,b)=-\beta(W^{M}(a,k,b)+\Delta f^{M}(k))+H(k),
\eeq
where
\beq
W^{M}(a,k,b)=E^{M}_{0}(a_{2})-E^{M}_{k}(b) \label{workcostmemory}
\eeq
is the work done by the memory, $\Delta f^{M}(k)=F^{M}_{k}-F^{M}_{0}$ is the free-energy difference, and $H(k)=-\ln p_{k}$ is the (unaveraged) Shannon entropy. We can derive the following quantum Jarzynski equality for the memory by using Eq.~(\ref{QMfluc}):
\beq
\aav{\ee^{\beta(W^{M}+\Delta f^{M})-H+I}}=1-\lambda^{\mathrm{meas}}. \label{QMGeneJar}
\eeq
Using the Jensen's inequality, Eq.~(\ref{QMGeneJar}) reproduces the generalized second law for the memory:
\beq
\aav{W^{M}}\leq \ -\aav{\Delta f^{M}}-\bz T(\av{I}-\av{H})+\bz T\ln(1-\lambda^{\mathrm{meas}}), \label{QMseclawab}
\eeq
where $\av{W^{M}}=\Tr[\rho^{M}_{0}H^{M}_{0}]-\sum_{k}p_{k}\Tr[\rho^{M}_{k}H^{M}_{k}]$ is the averaged work done by the memory and $\av{H}=-\sum_{k}p_{k}\ln p_{k}$ is the Shannon entropy. Note that $\av{W^{M}}$ usually takes a negative value since we need to input energy to the memory to perform the measurement. A nonzero $\lambda^{\mathrm{meas}}$ increases the work cost of the measurement due to absolute irreversibility as shown in Eq.~(\ref{QMseclawab}). Using the setup of our Hamiltonian of the memory in this section, $\lambda^{\mathrm{meas}}$ can be expressed as
\beq
\lambda^{\mathrm{meas}}=\sum_{x,a_{1}\neq 0,a_{2},k,y,b}p_{k}p^{S}(y|k)p^{M}_{\text{can}}(b|k)|\bbra{\psi^{S}(x)}\otimes \bbra{\phi^{M}_{a_{1}}(a_{2})} U^{\dagger SM}\kket{\varphi^{S}_{k}(y)}\otimes\kket{\phi^{M}_{k}(b)}|^{2} \label{Jarlambdam},
\eeq
where $p^{M}_{\mathrm{can}}(b|k)=\exp[-\beta(E^{M}_{k}(b)-F^{M}_{k})]$ is the canonical distribution corresponding to the initial state of the backward process, and $A=\{(x,a)|a=(0,a_{2})\}$ since $p^{S}_{\mathrm{can}}(x)\neq 0$ for all $x$ in this setup. From Eq.~(\ref{Jarlambdam}), we note that $\lambda^{\mathrm{meas}}$ is the total probability that the backward process ends in the subspace $\{\kket{\phi^{M}_{a_{1}}(a_{2}})\}_{a_{1}\neq 0,a_{2}}$, which was not occupied by the initial local equilibrium state~(\ref{initialcanom}). Note that the projection on the memory destroys the coherence between the system and the memory, which an irreversible process, and occurs only in the quantum regime due to the measurement back action.

\subsection{Quantum fluctuation theorems for feedback-controlled systems and unavailable information}
In this section, we consider the effect of absolute irreversibility during the feedback process in more detail. Without absolute irreversibility, the extra work beyond the conventional second law of thermodynamics that can be extracted from the system is bounded from above by $k_{\mathrm{B}}T$ times the obtained information $\av{I}$. However, if the feedback process is absolutely irreversible, we cannot fully utilize the information to extract work. We introduce the amount of information that is unavailable for use in extracting work for a given feedback control protocol as 
\beq
I_{\text{u}}(k)=-\ln(1-\lambda^{\mathrm{fb}}(k)). \label{unavailable}
\eeq
The quantity was originally introduced in Ref.~\cite{Ashida} for classical systems. Here, we introduce the total probability of the reference probability that does not go back to the postmeasurement state conditioned on the measurement outcome $k$:
\beq
\lambda^{\mathrm{fb}}(k)=\frac{1}{p_{k}}\sum_{h,k,y\not\in Y,j,z}p_{\mathrm{r}}^{\mathrm{fb}}(h,k,y,j,z). \label{unalamda}
\eeq
Then we start from the following relation:
\beq
p_{k}=\sum_{h,y\not\in Y,j,z}p^{\mathrm{fb}}(h,k,y,j,z)+\sum_{h,y\in Y,j,z}p^{\mathrm{fb}}(h,k,y,j,z). \label{pkrefprob}
\eeq
From Eqs.~(\ref{unavailable}) and (\ref{unalamda}), we find that Eq.~(\ref{pkrefprob}) takes the form
\beqa
\sum_{h,y\in Y,j,z}p^{\mathrm{fb}}(h,k,y,j,z)&=&p_{k}-p_{k}\cdot \lambda^{\mathrm{fb}}(k) \nonumber \\
&=&p_{k}\cdot \ee^{I_{\text{u}}(k)} . \label{sumhiu}
\eeqa
Multiplying both sides of Eq.~(\ref{sumhiu}) by $\ee^{-I_{\text{u}}(k)}$, summing over $k$ and using Eq.~(\ref{sigmaplusi}), we obtain
\beq
\aav{ \ee^{-\sigma^{SB}-(I-I_{\text{u}})} }=1.
\eeq
Using Jensen's inequality, we obtain the inequality for $\sigma^{SB}$ in the presence of unavailable information:
\beq
\aav{\sigma^{SB}}\geq -(\av{I}-\av{I_{\mathrm{u}}}). \label{inequnav}
\eeq
If we use the same assumptions~(\ref{Jarassuma}) and (\ref{Jarassumb}) in deriving the quantum Jarzynski equality, we obtain
\beq
\aav{ \ee^{\beta(W^{S}+\Delta f^{S})-(I-I_{\text{u}})} }=1,
\eeq
and hence
\beq
\aav{W^{S}}\leq -\aav{\Delta f^{S}}+k_{\mathrm{B}}T (\av{I}-\av{I_{\text{u}}}). \label{ineqwsav}
\eeq
The obtained inequalities~(\ref{inequnav}) and (\ref{ineqwsav}) give bounds on the entropy reduction of $SB$ and extractable work from the system, where they take into account the inefficiency of the feedback control by subtracting the unavailable information $\aav{I_{\mathrm{u}}}$ from the obtained information $\av{I}$. From the convexity, the unavailable information is bounded from above by
\beq
-\av{I_{\text{u}}}=-\sum_{k}p_{k}\ln(1-\lambda^{\mathrm{fb}}(k))\leq -\ln(1-\sum_{k}p_{k}\lambda^{\mathrm{fb}}(k))=-\ln(1-\lambda^{\mathrm{fb}}),
\eeq
so that inequality~(\ref{inequnav}) gives a tighter bound compared with inequality~(\ref{SBineqfb}).

\section{\label{sec:example} Examples}
\subsection{Quantum piston}
\begin{figure}[h,t]
\begin{center}
\includegraphics[width=\textwidth]{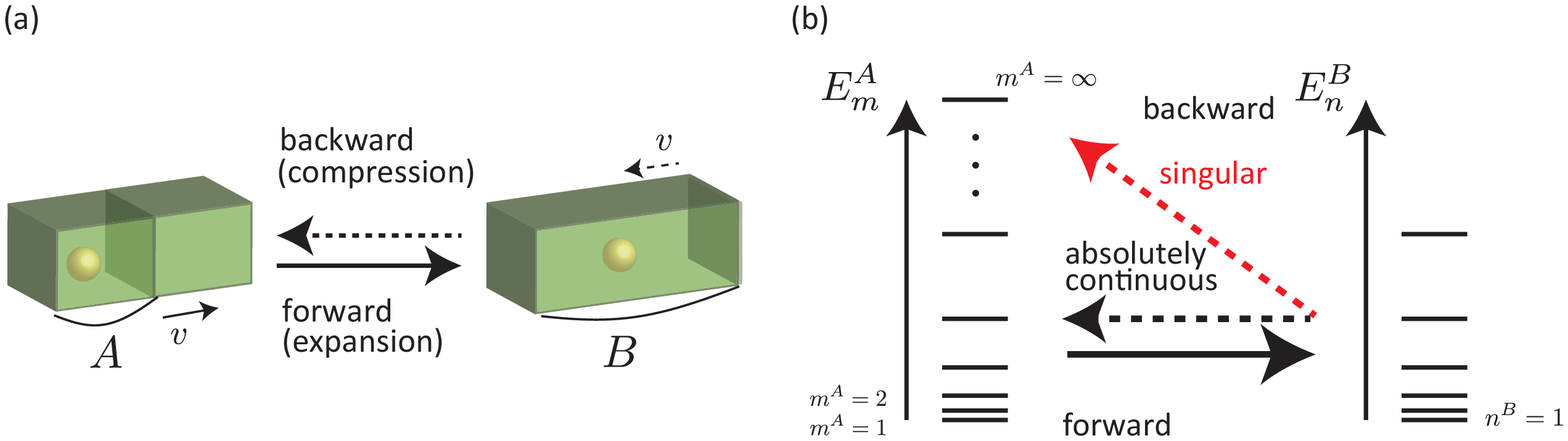}
\caption{\label{fig:qpiston} (a) Schematic illustration of the quantum piston model. A quantum particle is trapped inside a one-dimensional box. {\it Forward process} (expansion of the box): the initial length of the box is given by $A$ and the piston is pulled with a constant speed $v$. At the final time, the length of the box is given by $B$. {\it Backward process} (compression of the box): the initial lenght of the box is given by $B$ and the piston is pushed with a constant speed $v$ until the length of the box is given by $A$. (b) Schematic illustration of the transitions between the initial and final energy during the forward and backward protocols. We identify the backward path which ends up in the energy eigenstate $\kket{m^{A}=\infty}$ as a singular path.}
\end{center}
\end{figure}
In this subsection, we consider a free expansion of a gas by using a model of the quantum piston discussed in Refs.~\cite{Quan,Dodonov}, and apply the quantum Jarzynski equality~(\ref{Jarflucresult}) to this model.

We consider a quantum particle with mass $M$ trapped inside a one-dimensional box whose length is given by $L_{t}$ at time $t$. We start with $L_{0}=A$ and pull the piston with a constant speed $v$. At the final time, the length of the box is given by $L_{t_{\mathrm{f}}}=B>A$, meaning that $L_{t}=vt+A$ with $t_{\mathrm{f}}=(B-A)/v$. See Fig.~\ref{fig:qpiston}. The instantaneous energy eigenstate is denoted by $\ket{m^{L}}$, and its wave function is given by
\beq
\phi_{m}(x;L)=\sqrt{\frac{2}{L}}\sin\left(\frac{m\pi x}{L}\right),
\eeq
and the corresponding eigenenergy is given by
\beq
E^{L}_{m}=\frac{m^{2}\pi^{2}\hbar^{2}}{2ML^{2}}.
\eeq
We consider a process starting with a canonical distribution
\beq
\rho_{\mathrm{ini}}=\sum_{m}p_{\mathrm{ini}}(m^{A})\kket{m^{A}}\bbra{m^{A}},
\eeq
where
\beq
p_{\mathrm{ini}}(m^{A})=\frac{1}{Z_{A}}\ee^{-\beta E^{A}_{m}},
\eeq
with $Z_{A}=\sum_{m}\exp(-\beta E^{A}_{m})$ being the partition function. The conditional probability distribution of observing the initial state $\kket{m^{A}}$ and the final state $\kket{n^{B}}$ takes the form
\beq
p(n^{B}|m^{A})=|\bbra{n^{B}}U\kket{m^{A}}|^{2},
\eeq
where $U$ is the unitary operator describing the time evolution of the system. An explicit form of $p(n^{B}|m^{A})$ is given in Ref.~\cite{Quan} as
\beq
p(n^{B}|m^{A})=\left|\sum_{l=1}^{\infty}\frac{2}{A}\int^{A}_{0}\ee^{- \frac{\ii Mvx^{2}}{2A\hbar}}\sin\left(\frac{l\pi x}{A}\right)\sin\left(\frac{m\pi x}{A}\right)\dd x\  \ee^{-\frac{\ii\pi^{2}l^{2}\hbar (B-A)}{2ABMv}}\frac{2}{B}\int^{B}_{0}\ee^{ \frac{\ii Mvy^{2}}{2B\hbar}}\sin\left(\frac{l\pi y}{B}\right)\sin\left(\frac{n\pi y}{B}\right)\dd y \right|^{2}. \label{pistontrans}
\eeq
The forward probability distribution is then given by
\beq
p(n^{B},m^{A})=p_{\mathrm{ini}}(m^{A})p(n^{B}|m^{A}).
\eeq
Next, let us define the reference probability distribution of the backward process. To derive Jarzynski equality, we take the initial state of the backward process as the canonical distribution:
\beq
p_{\mathrm{r}}(n^{B})=\frac{1}{Z_{B}}\ee^{-\beta E^{B}_{n}},
\eeq
and the reference probability distribution is given by
\beq
p_{\mathrm{r}}(m^{A},n^{B})=p_{\mathrm{r}}(n^{B})\tilde{p}(m^{A}|n^{B}),
\eeq
where 
\beq
\tilde{p}(m^{A}|n^{B})=|\bbra{m^{A}}U^{\dagger}\kket{n^{B}}|^{2}=p(n^{B}|m^{A}).
\eeq
We derive the quantum Jarzynski equality by using the fact that the reference probability distribution is normalized:
\beqa
1&=&\sum_{m,n}p_{\mathrm{r}}(m^{A},n^{B}) \nonumber \\
&=&\sum_{m\not\in X,n}p_{\mathrm{r}}(m^{A},n^{B})+\sum_{m\in X,n}p_{\mathrm{r}}(m^{A},n^{B}), \label{pistonabs}
\eeqa
where $X$ is a set of $m^{A}$'s satisfying  $p_{\mathrm{r}}(m^{A},n^{B}) = 0$ if $p(m^{A},n^{B}) = 0$, specifying ordinary irreversible processes. We note that the term with $n^{B}=\infty$ does not contribute to the sum in Eq.~(\ref{pistonabs}) because $p_{\mathrm{r}}(m^{A},n^{B}=\infty)=0$. By taking the ratio between the forward probability and the backward probability, the second term on the right-hand side of Eq.~(\ref{pistonabs}) is related to the ensemble average of the exponentiated work:
\beqa
\sum_{m\in X,n}p_{\mathrm{r}}(m^{A},n^{B})&=&\sum_{m\in X,n}\frac{p_{\mathrm{r}}(m^{A},n^{B})}{p(m^{A},n^{B})}p(m^{A},n^{B}) \nonumber \\
&=&\sum_{m\in X,n}\ee^{\beta(E^{A}_{m}-E^{B}_{n}+F^{B}-F^{A})}p(m^{A},n^{B}) \nonumber \\
&=&\aav{\ee^{\beta(W+\Delta F)}},
\eeqa
where $W=E^{A}_{m}-E^{B}_{n}$ and $\Delta F=F^{B}-F^{A}$. By denoting the total probability of the absolutely irreversible process as $\lambda=\sum_{m\not\in X,n}p_{\mathrm{r}}(m^{A},n^{B})$, we can derive the quantum Jarzynski equality for a quantum piston:
\beq
\aav{\ee^{\beta(W+\Delta F)}}=1-\lambda. \label{pistonJareq}
\eeq

Next, we will show that for finite $v$, $\lambda=0$ as discussed in Ref.~\cite{Quan}. However, if we take the limit of $v=\infty$, $\lambda$ takes a nonzero value and the process is absolutely irreversible. The difference between the result given in Ref.~\cite{Quan} and ours is discussed in the appendix.

\begin{figure}[h,t]
\begin{center}
\includegraphics[width=.5\textwidth]{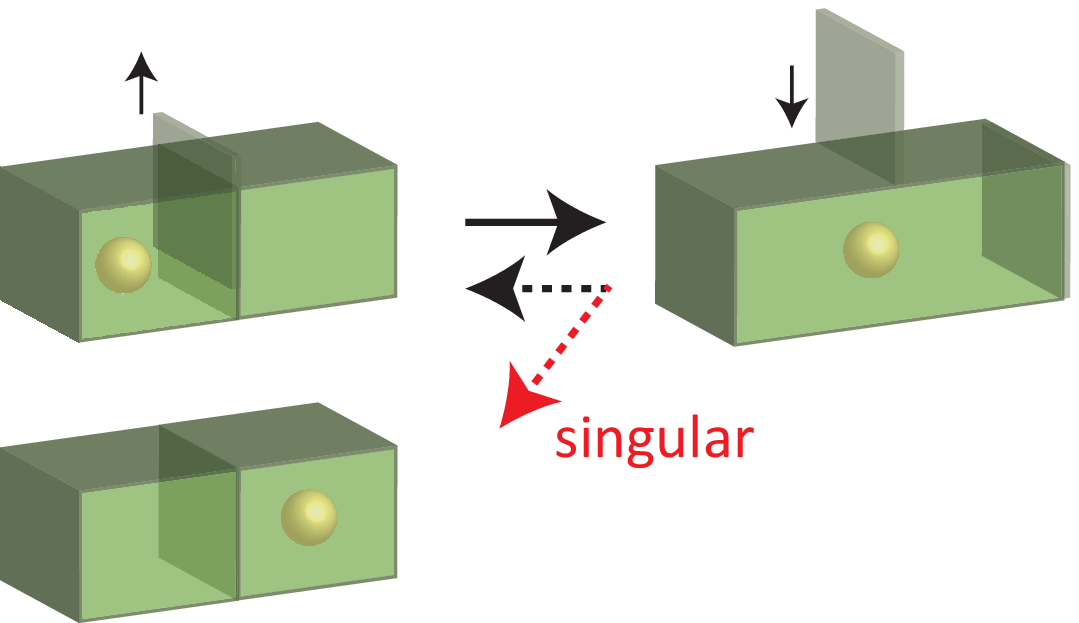}
\caption{\label{fig:removal} Schematic illustration of a free expansion of a gas. A quantum particle is trapped inside a one-dimensional box. The forward process is given by a sudden removal of the wall. The backward process is given by a sudden insertion of the wall. The particle ends up in either the left or right box, where the latter case is identified as a singular path. }
\end{center}
\end{figure}

Note that the forward probability vanishes only when the initial probability distribution $p_{\mathrm{ini}}(m^{A})$ vanishes. This occurs only for $m^{A}=\infty$, meaning that $X$ is given by a set of labels except $m^{A}=\infty$. (The contribution of $m^{A}=\infty$ is essential when we consider the case of $v=\infty$.) Then, $\lambda$ is given by
\beqa
\lambda&=&\sum_{n}p_{\mathrm{r}}(m^{A}=\infty,n^{B}) \nonumber \\
&=&\sum_{n}p_{\mathrm{r}}(n^{B})\tilde{p}(m^{A}=\infty|n^{B}). 
\eeqa

Let us first consider the case for finite $v$, and show that there is no absolute irreversibility in this case. By looking at Eq.~(\ref{pistontrans}), the integral over $x$ gives a nonzero value only when $l$ is of the order of $m$; otherwise the integrand oscillates rapidly due to the very large value of $m^{A}$. However, even if $l$ is large, the integral over $y$ vanishes due to the rapidly oscillating term, since $n^{B}$ takes a finite value. In conclusion, we have $\tilde{p}(m^{A}=\infty|n^{B})=0$ and
\beq
\lambda=0 \hspace{5mm} \text{for finite }v.
\eeq

Now we move on to the interesting case of $v=\infty$ (see Fig.~\ref{fig:qpiston}). We first consider the case in which $m^{A}$ is small compared with $v$ (meaning that in the limit $v\rightarrow \infty$, $m^{A}\in X$). By using the stationary phase approximation, $p(n|m)$ is given by (see Appendix for the detailed calculation)
\beq
p(n^{B}|m^{A})=\left|\int_{0}^{A}\frac{2}{\sqrt{AB}}\sin\left(\frac{m\pi z}{A}\right)\sin\left(\frac{n\pi z}{B}\right) \dd z\right|^{2}. \label{translow}
\eeq
For $m^{A}=\infty$ (or $m^{A}\not\in X$), the conditional probability does not vanish since the oscillatory term $\sin(m\pi x/A)$ cancels with the term $\exp(-\ii Mvx^{2}/(2A\hbar))$, and the conditional probability has a nonzero value: 
\beq
\lim_{v\rightarrow\infty}p(n^{B}|m^{A}=\infty)\neq 0.
\eeq
We refer to Ref.~\cite{Dodonov} for the direct calculation of $p(n^{B}|m^{A}=\infty)$. Instead, we use the following relation to calculate $\lambda$:
\beq
\sum_{m}\tilde{p}(m^{A}|n^{B})=1,
\eeq
which results from the unitary dynamics of the backward process. By using Eq.~(\ref{translow}), we obtain
\beqa
1-\tilde{p}(m^{A}=\infty|n^{B})&=&\sum_{m\in X}p(m^{A}|n^{B}) \nonumber \\
&=&\frac{2}{B}\int^{A}_{0}\dd x \sin^{2}\left(\frac{n\pi x}{B}\right) \nonumber \\
&=&\frac{A}{B}-\frac{\sin\left(\frac{2\pi n A}{B}\right)}{2\pi n},
\eeqa
and $\lambda$ is given by
\beqa
\lambda&=&\sum_{n}p_{\mathrm{r}}(n^{B})\tilde{p}(m^{A}=\infty|n^{B}) \nonumber \\
&=&\sum_{n}p_{\mathrm{r}}(n^{B})\left(1-\frac{A}{B}-\frac{\sin\left(\frac{2\pi n A}{B}\right)}{2\pi n}\right)\nonumber \\
&=&1-\frac{A}{B}-\sum_{n}\frac{\exp(-\frac{\beta n^{2}\pi^{2}\hbar^{2}}{2MA^{2}})}{Z^{B}}\frac{\sin\left(\frac{2\pi n A}{B}\right)}{2\pi n}. \label{pistonvlambda}
\eeqa

We also note that when we consider a sudden removal of the wall at $t=0$ instead of pulling the wall (as illustrated in Fig.~\ref{fig:removal}), the wave function of the initial state does not change. In this case, the exact form of the transition probability can be calculated by using the sudden approximation:
\beqa
\lim_{v\rightarrow \infty}U\kket{m^{A}}&=&\kket{m^{A}}, \\
p_{\text{sudden}}(n^{B}|m^{A})&=&|\braket{n^{B}}{m^{A}}|^{2} \nonumber \\
&=&\left|\int_{0}^{A}\frac{2}{\sqrt{AB}}\sin\left(\frac{m\pi x}{A}\right)\sin\left(\frac{n\pi x}{B}\right) \dd x\right|^{2}, \label{suddenlim}
\eeqa
which is equal to the transition probability for $m\in X$ in Eq.~(\ref{translow}). Therefore, the quantum Jarzynski equality for this process takes the same form as in Eq.~(\ref{pistonJareq}) with $\lambda$ given by Eq.~(\ref{pistonvlambda}).

Let us discuss the physical interpretation of the results. For $v\rightarrow \infty$ limit, the gas particle does not bounce at the wall in the forward process because the initial probability of having infinite energy $(m^{A}=\infty)$ is zero. On the other hand, if we consider a reverse process describing a infinitely fast compression of a box, the particle located in the region $A<x<B$ will be pushed by the wall and the final energy of the particle is given by $E^{A}_{m=\infty}$. Since the state $\ket{m^{A}=\infty}$ is not occupied in the initial state, such paths are absolutely irreversible and the total probability of those backward paths is given by $\lambda$. If we model the free expansion by a sudden removal of the wall, the forward process is exactly the same as the case of pulling the piston at infinite speed. However, the backward process is described by the sudden insertion of the wall, which is different from the sudden compression of the piston. In this case, the absolutely irreversible process is given by the paths where the particle ends up in the region $A<x<B$, which was not occupied in the initial state. In these two cases, $\lambda$ takes on the same value because they give the total probability of the backward process in which the particle is initially located in the region $A<x<B$.

\subsection{Feedback control on qubit systems}
\begin{figure}[h,t]
\begin{center}
\includegraphics[width=\textwidth]{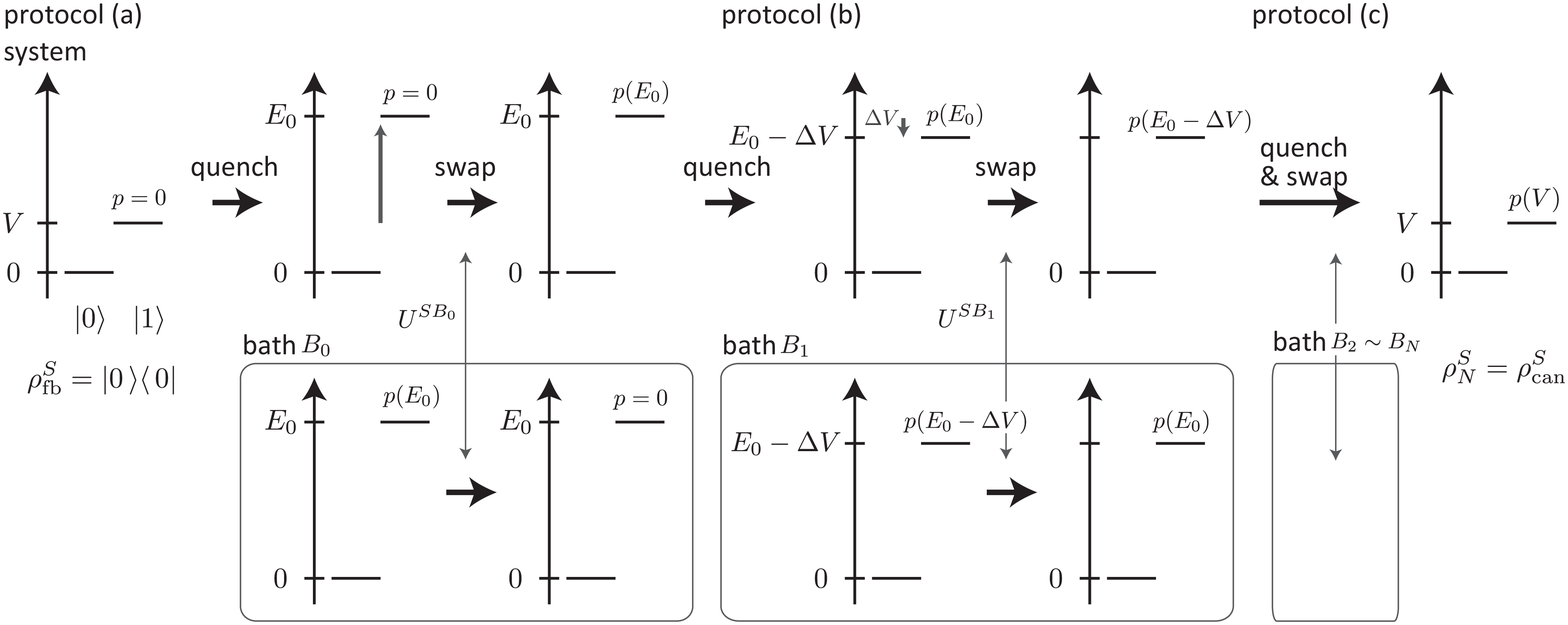}
\caption{\label{fig:qubit} Schematic illustration of the thermalization protocol. We consider a system composed of a qubit, and $N+1$ different heat baths $B_{0},\cdots,B_{N}$, each composed of a qubit. Here $p(E)=(1+\exp(\beta E))^{-1}$ denotes the the occupation probability of the state $\ket{1}$, which means that, the density matrix is given by the canonical distribution $\rho_{\mathrm{can}}(E):=(1-p(E))\ket{0}\bra{0}+p(E)\ket{1}\bra{1}$. The feedback control brings the state of the system to a pure state $\rho^{S}_{\mathrm{fb}}=\ket{0}\bra{0}_{S}$. We consider the following protocols (a)-(c) that transforms a pure state into the canonical distribution $\rho^{S}_{\mathrm{can}}(V)$. Protocol (a): We quench the energy level of the state $\ket{1}_{S}$ to $E_{0}$. Next, we prepare a heat bath in the canonical distribution with the energy level $E_{0}$. We swap the density matrices $\rho^{S}_{\mathrm{fb}}$ and $\rho^{B_{0}}_{\mathrm{can}}(E_{0})$ by applying $U^{SB_{0}}$, where energy is transfered from $B_{0}$ to $S$ during this process. After the swap, the density matrix of the system is given by $\rho^{S}_{\mathrm{can}}(E_{0})$. Protocol (b): We quench the system and lower the energy level by $\Delta V$, and energy is extracted from the system. We prepare a heat bath $B_{1}$ in the canonical distribution $\rho^{B}_{\mathrm{can}}(E_{0}-\Delta V)$ and swap the density matrices between $S$ and $B_{1}$, where energy is transfered from $B_{1}$ to $S$. After the swap, the density matrix of the system is given by $\rho^{S}_{\mathrm{can}}(E_{0}-\Delta V)$. Protocol (c): We repeat the protocol which is similar to the protocol (b) by lowering the energy level by $\Delta V$ (quench) and swapping the density matrices between $S$ and $B_{n}$ ($2\leq n\leq N$) and the density matrix of the system is given by $\rho^{S}_{\mathrm{can}}(E_{0}-n\Delta V)$. After the $N$th protocol, the density matrix of the system is transformed into the canonical distribution $\rho^{S}_{\mathrm{can}}(V)$ and the energy level of the system is returned to $V$, which completes the thermalization process.}
\end{center}
\end{figure}
In this subsection, we apply the quantum Jarzynski equality~(\ref{QCGeneJar}) to qubit systems. Let us prepare an initial state given by
\beq
\rho^{S}_{\mathrm{ini}}=\rho^{S}_{\mathrm{can}}=\frac{1}{1+\ee^{-\beta V}}\ket{0}\bra{0}_{S}+\frac{\ee^{-\beta V}}{1+\ee^{-\beta V}}\ket{1}\bra{1}_{S},
\eeq
where $V$ is the energy difference between the two states $\ket{0}_{S}$ and $\ket{1}_{S}$.

Let us perform a projective measurement with respect to the basis set $\{\ket{0}_{S},\ket{1}_{S}\}$. The  probability $p_{k}$ of obtaining the measurement outcome $k$ is given by 
\beqa
p_{0}&=&\frac{1}{1+\ee^{-\beta V}}  \\
p_{1}&=&\frac{\ee^{-\beta V}}{1+\ee^{-\beta V}},
\eeqa
and the postmeasurement state conditioned on the measurement outcome is given by a pure state
\beq
\rho^{S}(k)=\kket{k}\bbra{k}_{S}.
\eeq
We can confirm that the acquired knowledge of the system $\av{I}$ is equal to the Shannon entropy $H(p_{k})=-\sum_{k}p_{k}\ln p_{k}$ calculated from the probability distribution of the measurement outcome:
\beqa
\av{I}&=&S(\rho^{S}_{\mathrm{ini}})-\sum_{k}p_{k}S(\rho^{S}(k)) \nonumber \\
&=&H(p_{k})=\beta V\cdot\frac{\ee^{-\beta V}}{1+\ee^{-\beta V}}+\ln(1+\ee^{-\beta V}).
\eeqa
Note that the state of the system conditioned on the measurement outcome is less mixed for a greater value of $\av{I}$. In this case, the postmeasurement state is given by a pure state and our knowledge of the state has increased by the maximum amount $\av{I}=H(p_{k})$ due to the measurement.

Depending on the measurement outcome, we perform the following feedback control which has the effect of flipping the state if the post-measurement state is $\ket{1}_{S}$:
\beq
U_{0}=1, \ U_{1}=\ket{0}\bra{1}_{S}+\ket{1}\bra{0}_{S}. 
\eeq
After the feedback control, we obtain
\beq
\rho^{S}_{\mathrm{fb}}(k)=U_{k}\ket{k}\bra{k}U^{\dagger}_{k}=\ket{0}\bra{0}_{S}, \label{ex:fb}
\eeq
which is independent of the measurement outcome $k$. The averaged density matrix is given by $\rho^{S}_{\mathrm{fb}}=\sum_{k}p_{k}\rho^{S}_{\mathrm{fb}}(k)=\ket{0}\bra{0}_{S}$. The energy change of the system during the feedback is given by
\beq
\Delta E_{\mathrm{fb}}=-V\frac{\ee^{-\beta V}}{1+\ee^{-\beta V}}.
\eeq

We model the thermalization process by introducing $N+1$ different heat baths, each of which is composed of a qubit, as schematically illustrated in Fig.~\ref{fig:qubit}. A similar model is discussed in Ref.~\cite{qubitext}. The Hamiltonian of each heat bath is given by
\beq
H^{B_{n}}=(E_{0}-n\Delta V )\ket{1}\bra{1}_{B_{n}}, \hspace{5mm} n=0,\cdots N,
\eeq
where $E_{0}=N\Delta V+V$ is the energy difference between two states of the zeroth heat bath. The initial state of the entire heat bath is given by the tensor product of the canonical distributions:
\beq
\rho_{\mathrm{can}}^{B}=\bigotimes_{n}\left(\frac{1}{1+\ee^{-\beta(E_{0}-n\Delta V)}}\ket{0}\bra{0}_{B_{n}}+\frac{\ee^{-\beta(E_{0}-n\Delta V)}}{1+\ee^{-\beta(E_{0}-n\Delta V)}}\ket{1}\bra{1}_{B_{n}}\right).
\eeq
We consider the following $N+1$ steps of the protocol to thermalize the system.

(a) We quench the Hamiltonian of the system so that the energy difference of the system is changed from $E$ to $E_{0}$. Note that this process preserves the energy of the system since the excited state $\ket{1}_{S}$ is not populated during this process. Next, we perform the following unitary transformation between the system and $B_{0}$:
\beq
U^{SB_{0}}=\ket{0}\bra{0}_{S}\otimes\ket{0}\bra{0}_{B_{0}}+\ket{1}\bra{0}_{S}\otimes\ket{0}\bra{1}_{B_{0}}+\ket{0}\bra{1}_{S}\otimes\ket{1}\bra{0}_{B_{0}}+\ket{1}\bra{1}_{S}\otimes\ket{1}\bra{1}_{B_{0}}. \label{unitaryfeedbackex}
\eeq
This swaps the populations between $S$ and $B_{0}$:
\beqa
\rho^{SB_{0}}_{0}&=&U^{SB_{0}}\left(\ket{0}\bra{0}_{S}\otimes\left(\frac{1}{1+\ee^{-\beta E_{0}}}\ket{0}\bra{0}_{B_{0}}+\frac{\ee^{-\beta E_{0}}}{1+\ee^{-\beta E_{0}}}\ket{1}\bra{1}_{B_{0}}\right)\right)U^{\dagger SB_{0}} \nonumber \\
&=& \left(\frac{1}{1+\ee^{-\beta E_{0}}}\ket{0}\bra{0}_{S}+\frac{\ee^{-\beta E_{0}}}{1+\ee^{-\beta E_{0}}}\ket{1}\bra{1}_{S}\right)\otimes \ket{0}\bra{0}_{B_{0}}.
\eeqa
During this process, the energy flow occurs from $B_{0}$ to $S$. The energy change $\Delta E=E_{\mathrm{fin}}-E_{\mathrm{ini}}$ of $B_{0}$ and that of $S$ can be explicitly calculated as
\beqa
\Delta E^{B_{0}}&=&-E_{0}\cdot\frac{\ee^{-\beta E_{0}}}{1+\ee^{-\beta E_{0}}}, \label{heatzero} \\
\Delta E^{S}_{0}&=&E_{0}\cdot\frac{\ee^{-\beta E_{0}}}{1+\ee^{-\beta E_{0}}}.
\eeqa
We note that the total energy change is zero: $\Delta E^{B_{0}}+\Delta E^{S}_{0}=0$.

(b) We quench the Hamiltonian of the system so that the energy difference is changed from $E_{0}$ to $E_{0}-\Delta V$. During this process, the energy change of the system is given by
\beq
\Delta E_{\mathrm{q},1}^{S}=-\Delta V\cdot \frac{\ee^{-\beta E_{0}}}{1+\ee^{-\beta E_{0}}} .
\eeq
Next, we let $S$ interact with $B_{1}$ via the unitary transformation which has the same form of Eq.~(\ref{unitaryfeedbackex}). After the swap, the density matrix is given by
\beqa
\rho^{SB_{1}}_{1}&=&U^{SB_{1}}(\rho^{S}_{0}\otimes\rho^{B_{1}}_{\mathrm{can}})U^{\dagger SB_{1}} \nonumber \\
&=& \left(\frac{1}{1+\ee^{-\beta (E_{0}-\Delta V)}}\ket{0}\bra{0}_{S}+\frac{\ee^{-\beta (E_{0}-\Delta V)}}{1+\ee^{-\beta (E_{0}-\Delta V)}}\ket{1}\bra{1}_{S}\right) \nonumber \\
& &\otimes \left(\frac{1}{1+\ee^{-\beta E_{0}}}\ket{0}\bra{0}_{B_{1}}+\frac{\ee^{-\beta E_{0}}}{1+\ee^{-\beta E_{0}}}\ket{1}\bra{1}_{B_{1}}\right),
\eeqa
and the energy changes of $B_{1}$ and $S$ are given by
\beq
\Delta E^{B_{1}}=-(E_{0}-\Delta V)\cdot\left( \frac{\ee^{-\beta (E_{0}-\Delta V)}}{1+\ee^{-\beta (E_{0}-\Delta V)}}-\frac{\ee^{-\beta E_{0}}}{1+\ee^{-\beta E_{0}}}  \right)=-\Delta E^{S}_{1}.
\eeq

(c) For the $n$th step ($2\leq n\leq N$), we quench the Hamiltonian of the system so that the energy difference is changed from $E_{0}-(n-1)\Delta V$ to $E_{0}-n\Delta V$. During this process, the energy change of the system is given by
\beq
\Delta E^{S}_{\mathrm{q},n}=-\Delta V  \frac{\ee^{-\beta (E_{0}-(n-1)\Delta V)}}{1+\ee^{-\beta (E_{0}-(n-1)\Delta V)}} .
\eeq
Next, we interact $S$ and $B_{n}$ using the unitary transformation which has the same form as Eq.~(\ref{unitaryfeedbackex}). After the swap, the density matrix is given by
\beqa
\rho^{SB_{n}}_{n}&=&U^{SB_{n}}(\rho^{S}_{n-1}\otimes\rho^{B_{n}}_{\mathrm{can}})U^{\dagger SB_{n}} \nonumber \\
&=& \left(\frac{1}{1+\ee^{-\beta (E_{0}-n\Delta V)}}\ket{0}\bra{0}_{S}+\frac{\ee^{-\beta (E_{0}-n\Delta V)}}{1+\ee^{-\beta (E_{0}-n\Delta V)}}\ket{1}\bra{1}_{S}\right) \nonumber \\
& &\otimes \left(\frac{1}{1+\ee^{-\beta (E_{0}-(n-1)\Delta V)}}\ket{0}\bra{0}_{B_{n}}+\frac{\ee^{-\beta (E_{0}-(n-1)\Delta V)}}{1+\ee^{-\beta (E_{0}-(n-1)\Delta V)}}\ket{1}\bra{1}_{B_{n}}\right),
\eeqa
and the energy change is given by
\beq
\Delta E^{B_{n}}=-(E_{0}-n\Delta V)\cdot\left(  \frac{\ee^{-\beta (E_{0}-n\Delta V)}}{1+\ee^{-\beta (E_{0}-n\Delta V)}}-\frac{\ee^{-\beta (E_{0}-(n-1)\Delta V)}}{1+\ee^{-\beta (E_{0}-(n-1)\Delta V)}} \right)=-\Delta E^{S}_{n}.
\eeq

After the $N$th step, the system returns to the canonical distribution, which is the final state of this protocol:
\beq
\rho^{S}_{N}=\rho^{S}_{\mathrm{can}}.
\eeq
We use the short-hand notation
\beq
\ket{h}_{B}=\ket{h_{0}}_{B_{0}}\otimes\ket{h_{1}}_{B_{1}}\otimes\cdots\otimes\ket{h_{N}}_{B_{N}}, 
\eeq
where $h_{n}$ takes the value $0$ or $1$, and $\ket{h_{n}}_{B_{n}}$ describes the energy eigenstate of the $n$th heat bath. We also use the notation $U^{SB}=U^{SB_{N}}U^{SB_{N-1}}\cdots U^{SB_{0}}$, which is the total unitary operation performed on the total system during the thermalization process. Now we explicitly calculate the left-hand side of Eq.~(\ref{QCGeneJar}):~\footnote{While we derive Eq.~(\ref{QCGeneJar}) under a single heat bath, a generalization to multiple heat baths is straightforward.}
\beq
\aav{\ee^{-\beta W-I}}=\sum_{h,j,k,z}\left|\bra{z}_{S}\otimes\bra{j}_{B}U^{SB}U^{S}_{k}\ket{k}_{S}\otimes\ket{h}_{B}\right|^{2}\frac{\ee^{-\beta E^{B}(h)}}{Z_{B}}\frac{\ee^{-\beta E^{S}(k)}}{Z_{S}}\ee^{\beta W(k,z,h,j)-I(k)}, \label{ex:genejar}
\eeq
where
\beq
\frac{\ee^{-\beta E^{B}(h)}}{Z_{B}}=\frac{\ee^{-\beta E^{B_{0}}(h_{0})}}{Z_{B_{0}}}\cdot\frac{\ee^{-\beta E^{B_{1}}(h_{1})}}{Z_{B_{1}}}\cdots\frac{\ee^{-\beta E^{B_{N}}(h_{N})}}{Z_{B_{N}}},
\eeq
and
\beqa
E^{B_{n}}(h_{n}) &=&\begin{cases} 0 & \hspace{4mm} h_{n}=0, \\ E_{0}-n\Delta V & \hspace{4mm} h_{n}=1,     \end{cases} \\
E^{S}(k)&=&\begin{cases} 0 & \hspace{18mm} k=0, \\ V & \hspace{18mm} k=1.     \end{cases}
\eeqa
Noting that the work in Eq.~(\ref{ex:genejar}) is given by
\beq
W(h,j,k,z)=E^{S}(k)-E^{S}(z)+E^{B}(h)-E^{B}(j),
\eeq
we can further calculate Eq.~(\ref{ex:genejar}) and obtain
\beqa
\aav{\ee^{-\beta W-I}}&=&\sum_{h,j,k,z}p_{k}\left|\bra{z}_{S}\otimes\bra{j}_{B}U^{SB}U^{S}_{k}\ket{k}_{S}\otimes\ket{h}_{B}\right|^{2}\frac{e^{-\beta E^{B}(j)}}{Z_{B}}\frac{e^{-\beta E^{S}(z)}}{Z_{S}} \nonumber \\.
&=&\sum_{k}p_{k}\bra{k}_{S}U^{\dagger S}_{k} \Tr_{B}[U^{\dagger SB}_{k}(\rho^{S}_{\mathrm{can}}\otimes\rho^{B}_{\mathrm{can}})U^{SB}]  U^{S}_{k}\ket{k}_{S}.
\eeqa
Due to the reverse protocol given above, the density matrix of the system returns to the state
\beq
\Tr_{B}[U^{\dagger SB}(\rho^{S}_{\mathrm{can}}\otimes\rho^{B}_{\mathrm{can}})U^{SB}] =\frac{1}{1+\ee^{-\beta E_{0}}}\ket{0}\bra{0}_{S}+\frac{\ee^{-\beta E_{0}}}{1+\ee^{-\beta E_{0}}}\ket{1}\bra{1}_{S}. \label{ex:reverse}
\eeq
Using Eq.~(\ref{ex:reverse}), we have an explicit form of Eq.~(\ref{ex:genejar}):
\beq
\aav{\ee^{-\beta W-I}}=\frac{1}{1+\ee^{-\beta E_{0}}}=1-\lambda^{\mathrm{fb}}, \label{ex:jarfbfin}
\eeq
where $\lambda^{\mathrm{fb}}$ is the total probability of the backward process not returning to the postmeasurement state $\ket{k}_{S}$ as given in Eq.~(\ref{projectionlambdafb}):
\beqa
\lambda^{\mathrm{fb}}&=&\sum_{k\neq y}p_{k}\bra{y}_{S}U^{\dagger S}_{k} \Tr_{B}[U^{\dagger SB}(\rho^{S}_{\mathrm{can}}\otimes\rho^{B}_{\mathrm{can}})U^{SB}]  U^{S}_{k}\ket{y}_{S} \nonumber \\
&=&\frac{\ee^{-\beta E_{0}}}{1+\ee^{-\beta E_{0}}}.
\eeqa
Using Jensen's inequality and Eq.~(\ref{ex:jarfbfin}), we can derive the upper bound on extractable work from the system via feedback control, that is,
\beq
\av{W}\leq k_{\mathrm{B}}T\av{I}+k_{\mathrm{B}}T\ln(1-\lambda^{\mathrm{fb}}). \label{ex:sl}
\eeq
The right-hand side of~(\ref{ex:sl}) can be explicitly calculated as
\beq
k_{\mathrm{B}}T(\av{I}+\ln(1-\lambda^{\mathrm{fb}}))=V\frac{\ee^{-\beta V}}{1+\ee^{-\beta V}}+k_{\mathrm{B}}T\ln\frac{1+\ee^{-\beta V}}{1+\ee^{-\beta E_{0}}}.
\eeq
We can also calculate the work defined in Eq.~(\ref{workfeedback}):
\beq
\av{W}=-\Delta E^{S}-\Delta E^{B}, \label{Jarworkdef}
\eeq
where $\Delta E^{S}$ and $\Delta E^{B}$ are the total energy change of $S$ and $B$, respectively. As the system returns to the initial state at the end of the protocol, $\Delta E^{S}=0$. The total energy change of the heat bath is given by
\beq
-\Delta E^{B}=-\sum_{n}\Delta E^{B_{n}}=\sum_{n=0}^{N-1}\Delta V \frac{\ee^{-\beta(E_{0}-n\Delta V)}}{1+\ee^{-\beta(E_{0}-n\Delta V)}}+V \frac{\ee^{-\beta V}}{1+\ee^{-\beta V}}. \label{ex:energychangeheat}
\eeq
We can also interpret work as the energy extraction during the quench process during (a)-(c) combined with the energy extraction during the flipping process of the feedback control, that is,
\beqa
\av{W}&=&-\sum_{n}\Delta E^{S}_{\mathrm{q},n}-\Delta E^{S}_{\mathrm{fb}} \nonumber \\
&=&\sum_{n=0}^{N-1}\Delta V \frac{\ee^{-\beta(E_{0}-n\Delta V)}}{1+\ee^{-\beta(E_{0}-n\Delta V)}}+V \frac{\ee^{-\beta V}}{1+\ee^{-\beta V}},
\eeqa
which gives the same amount of work compared with the extracted work defined in Eq.~(\ref{Jarworkdef}). As we fix $E_{0}=N\Delta V+V$ and take the limit $\Delta V\rightarrow 0$ (and $N\rightarrow \infty$), the right-hand side of Eq.~(\ref{ex:energychangeheat}) reaches
\beq
\Delta E^{B}|_{\Delta V\rightarrow 0}=\int^{E_{0}}_{0}\dd V\frac{\ee^{-\beta(E_{0}-V)}}{1+\ee^{-\beta(E_{0}-V)}}=V\cdot \frac{\ee^{-\beta V}}{1+\ee^{-\beta V}}+k_{\mathrm{B}}T\ln\frac{1+\ee^{-\beta V}}{1+\ee^{-\beta E_{0}}}. \label{ex:limit}
\eeq
Since $\Delta E^{B}\leq \Delta E^{B}|_{\Delta V\rightarrow 0}$, inequality~(\ref{ex:sl}) is valid and the equality condition is achieved  in the limit of $\Delta V\rightarrow 0$ and $N\rightarrow \infty$:
\beq
\av{W}=k_{\mathrm{B}}T\av{I}+k_{\mathrm{B}}T\ln(1-\lambda^{\mathrm{fb}}), \hspace{5mm} (\Delta V\rightarrow 0). \label{ex:eqachie}
\eeq
If we consider a finite $\Delta V$, the density matrix of $S$ jumps from $\rho^{S}_{n}$ to $\rho^{S}_{n+1}$ during the process, causing dissipation. This dissipation is due to the ordinary irreversibility of the process and not due to absolute irreversibility, since only the relative weights of two states $\ket{0}_{S}$ and $\ket{1}_{S}$ are changed during the protocols between (a) and (c). 
 
The effect of absolute irreversibility depends on the parameter $E_{0}$ for this model, since the protocol (a) brings the pure state  $\ket{0}_{S}$ of the postmeasurement state into a thermal state.  If we take the limit $E_{0}\rightarrow \infty$, we have no absolute irreversibility ($\lambda^{\mathrm{fb}}=0$) and the backward process corresponding to the forward protocol (a) makes the density matrix return to $\ket{0}_{S}$. In this limit, one can extract work up to the amount commensurate with information obtained via measurement:
\beq
\av{W}\leq k_{\mathrm{B}}T\av{I}\hspace{5mm}(E_{0}\rightarrow\infty),
\eeq
and the equality is achieved again in the $\Delta V\rightarrow 0$ limit:
\beq
\av{W}=k_{\mathrm{B}}T\av{I}\hspace{5mm} (\Delta V\rightarrow 0, \ E_{0}\rightarrow \infty), \label{ex:equality}
\eeq
where the acquired information is fully utilized to extract work. The protocol we consider (in the limit of $\Delta V\rightarrow 0$ and $E_{0}\rightarrow \infty$, so that the system interacts with infinitely many heat baths) gives a quasi-static process of the isothermal expansion of the system in the sense that 
\beq
\Delta S^{S}=\beta Q
\eeq
is achieved, where $\Delta S^{S}=S(\rho^{S}_{\mathrm{can}})-S(\ket{0}\bra{0}_{S})$ gives a change in the von Neumann entropy of the system during the thermalization process (a) - (c) and 
\beq
Q=\Delta E^{B}
\eeq
is the heat taken from the heat baths. We relate the energy change of the heat baths to heat because the total change in the von Neumann entropy of the heat bath satisfies the thermodynamic relation 
\beq
\Delta S^{B}=\sum_{n}S(\rho^{B_{n}}_{n})-S(\rho^{B}_{\mathrm{can}})=-\beta \Delta E^{B}.
\eeq
As a result, Eq.~(\ref{ex:equality}) is satisfied and the reduced entropy of the system via feedback is fully converted into work by this quasi-static process. This result is to be campared  with the classical single-particle Szilard engine that achieves Eq.~(\ref{ex:equality}) via quasistatic isothermal expansion of the box~\cite{Szilard}.

Next, let us consider the opposite limit of $E_{0}=V$. In this case, we do not quench the energy level of the system. We only attach a single heat bath, letting the postmeasurement state of the system transform into a thermal state by a single jump (the protocol (a)). The effect of absolute irreversibility is maximal in this limit: 
\beq
\lambda^{\mathrm{fb}}|_{E_{0}=V}=\frac{\ee^{-\beta V}}{1+\ee^{-\beta V}} \leq \frac{\ee^{-\beta E_{0}}}{1+\ee^{-\beta E_{0}}},
\eeq
and the work gain takes the smallest value
\beq
\av{W}_{E_{0}=V}=V\frac{\ee^{-\beta V}}{1+\ee^{-\beta V}}.
\eeq

\section{\label{sec:conclusion} Conclusion}
We have derived the quantum fluctuation theorem~(\ref{Qfluc}) and Jarzynski equality~(\ref{Jarflucresult}) in the presence of absolutely irreversible processes, where the density matrix of the backward process does not return to the subspace spanned by the eigenvectors that have nonzero weight of the initial density matrix. We have also derived equalities for feedback and measurement processes~(\ref{QCfluc}) and (\ref{QMfluc}). The effect of absolute irreversibility limits the work gain via (inefficient) feedback control and also gives additional entropy production due to the projection on the memory~(\ref{SBineqfb}) and (\ref{Mineqmeas}). The latter fact means that the dissipation by quantum decoherence can be qualitatively characterized by absolute irreversibility. We have also discussed a model of the quantum piston and the feedback control on a qubit system to illustrate the obtained nonequilibrium equalities. 

\section*{Acknowledgments}
This work was supported by KAKENHI Grant No. 26287088 from the Japan Society for the Promotion of Science, 
a Grant-in-Aid for Scientific Research on Innovation Areas ``Topological Quantum Phenomena'' (KAKENHI Grant No. 22103005),
the Photon Frontier Network Program from MEXT of Japan, and the Mitsubishi Foundation. K. F. acknowledges support from JSPS (Grant No. 254105) and through Advanced Leading Graduate Course for Photon Science (ALPS). Y.M. was supported by the Japan Society for the Promotion of Science through Program for Leading Graduate Schools (MERIT). K.F. thanks Adolfo del Campo, Jordan M. Horowitz and Yuto Ashida for fruitful discussions.

\appendix
\section{Quantum pistons}
\subsection{Derivation of Eq~(\ref{translow})}
We wish to derive Eq.~(\ref{translow}) for small $m$. Note that the integral over $x$ in Eq.~(\ref{pistontrans}) vanishes for small $l$ due to the rapidly oscillating terms. If $l$ is the order of $v$, the integral can be approximated by using the stationary phase approximation:
\beq
\frac{2}{A}\int^{A}_{0}\ee^{- \frac{\ii Mvx^{2}}{2A\hbar}}\sin\left(\frac{l\pi x}{A}\right)\sin\left(\frac{m\pi x}{A}\right)\dd x\simeq-\frac{\ii}{A}\ee^{\frac{\ii\pi^{2}l^{2}\hbar}{2AMv}}\sin\left(\frac{lm\pi^{2}\hbar}{AMv}\right)\sqrt{\frac{2\pi  A\hbar}{Mv}}\ee^{-\frac{\ii\pi}{4}},
\eeq
where $(l\pi\hbar)/Mv\leq A$. We can also approximate the integral for $y$:
\beq
\frac{2}{B}\int^{B}_{0}\ee^{ \frac{\ii Mvy^{2}}{2B\hbar}}\sin\left(\frac{l\pi y}{A}\right)\sin\left(\frac{n\pi y}{B}\right)\dd y\simeq \frac{\ii}{B}\ee^{-\frac{\ii\pi^{2}l^{2}\hbar}{2BMv}}\sin\left(\frac{ln\pi^{2}\hbar}{BMv}\right)\sqrt{\frac{2\pi B\hbar}{Mv}}\ee^{\frac{\ii\pi}{4}},
\eeq
where $(l\pi\hbar)/Mv\leq B$. Now $P(n|m)$ takes the form
\beq
p(n^{B}|m^{A})=\left|\sum_{l}\frac{\hbar}{Mv}\frac{2\pi}{\sqrt{AB}}\sin\left(\frac{lm\pi^{2}\hbar}{AMv}\right)\sin\left(\frac{ln\pi^{2}\hbar}{BMv}\right)\right|^{2}, \label{transsuml}
\eeq
where $(l\pi\hbar)/Mv\leq A\leq B$. Note that we extended the summation to small $l$, since they give a negligible contribution when we take the limit of $v\rightarrow \infty$. By taking the $v\rightarrow \infty$ limit, we replace the sum with the integral, i.e., 
\beq
z=\frac{l\pi\hbar}{Mv},\hspace{5mm} \dd z=\frac{\pi\hbar}{Mv}. \label{transsumr}
\eeq
Then, combining Eqs.~(\ref{transsuml}) and (\ref{transsumr}) gives the desired result~(\ref{translow}).

\subsection{Comparison with previous work}
In this subsection, we compare our results with the results derived in Ref.~\cite{Quan}. In Ref.~\cite{Quan}, the authors derived the quantum Jarzynski equality for some finite $v$ and obtained Eq.~(\ref{pistonJareq}) with $\lambda=0$. They also discussed the different order of taking the limits for the velocity $v$ and the number $N$ of samplings of the ensemble average. To be specific, let us write the left-hand side of the Jarzynski equality with a fixed velocity $v$ and take the ensemble average for some finite number $N$: 
\beq
f_{N,v}:=\av{\exp(\beta(W+\Delta F))}_{N,v}.
\eeq
The authors of Ref.~\cite{Quan} discussed the validity of the Jarzynski equality for a free expansion process by taking the limit $v$ to infinity after taking $N$ to infinity: $\lim_{v\rightarrow\infty}\lim_{N\rightarrow\infty}f_{N,v}=1$. However, this operation implicitly assumes that a particle has to bounce at the infinitely fast moving wall. Therefore, it is questionable to interpret this process as a free expansion. In the mathematical sense, this limit corresponds to taking the following limits:
\beqa
\lim_{v\rightarrow\infty}\lim_{N\rightarrow\infty}f_{N,v}&=&\lim_{v\rightarrow \infty}\sum_{m\in X,n}p_{\mathrm{r}}(m^{A},n^{B})+\lim_{v\rightarrow \infty}\sum_{m\not\in X,n}p_{\mathrm{ini}}(m^{A})\ee^{\beta(E^{A}_{m}-\Delta F^{A})}p_{\mathrm{r}}(m^{A},n^{B}) \nonumber \\
&=&\lim_{v\rightarrow \infty}\sum_{m\in X,n}p_{\mathrm{r}}(m^{A},n^{B})+\lim_{v\rightarrow \infty}\sum_{m\not\in M,n}\frac{p_{\mathrm{ini}}(m^{A})}{p_{\mathrm{ini}}(m^{A})}p_{\mathrm{r}}(m^{A},n^{B}). \label{Quanresult}
\eeqa
Here special care should be taken to the second term on the right-hand side of Eq.~(\ref{Quanresult}) because the $v\rightarrow \infty$ limit means
\beq
\lim_{v\rightarrow \infty}\sum_{m\not\in X,n}\frac{p_{\mathrm{ini}}(m^{A})}{p_{\mathrm{ini}}(m^{A})}p_{\mathrm{r}}(m^{A},n^{B})=\sum_{n}\frac{0}{0}\ p_{\mathrm{r}}(m^{A}=\infty,n^{B}).
\eeq
Only if we assume $0/0=1$, the right-hand side of Eq.~(\ref{Quanresult}) is equal to the identity and the quantum Jarzynski equality is obtained:
\beq
\lim_{v\rightarrow\infty}\lim_{N\rightarrow\infty}f_{N,v}=1. \label{limvN}
\eeq
On the other hand, our result in Eq.~(\ref{pistonJareq}) consider the other ordering of the limits:
\beq
\lim_{N\rightarrow\infty}\lim_{v\rightarrow\infty}f_{N,v}=1-\lambda,
\eeq
which describes a more physical situation of a free expansion compared with that in Eq.~(\ref{limvN}).

\end{document}